\documentclass[onecolumn,]{aastex631}

\submitjournal{ApJ}

\shorttitle{SSC CRs}
\shortauthors{Fraija, Dai notti, Sahil, Debarpita, Warren}

\usepackage{epstopdf}
\usepackage{ulem}
\usepackage{amssymb}
\usepackage{times}
\usepackage[usenames,dvipsnames]{pstricks}
\usepackage{psfrag}
\usepackage{lipsum}
\usepackage{natbib}
\usepackage{textcase}

\usepackage{booktabs}
\usepackage{comment}
\usepackage{graphicx,subfigure}
\usepackage{textgreek}
\usepackage{hyperref}
\def\apjl{ApJ}%% Astrophysical Journal, Letters
\def\apjs{ApJS}%% Astrophysical Journal, Supplement
\def\ao{Appl.~Opt.}%% Applied Optics
\def\aap{A\&A}%% Astronomy and Astrophysics
\newcommand{\be}{\begin{equation}}
\newcommand{\ee}{\end{equation}}
\newcommand{\bary}{\begin{eqnarray}}
\newcommand{\eary}{\end{eqnarray}}

\begin{document}

\title{Synchrotron Self-Compton Afterglow Closure Relations and Fermi-LAT Detected Gamma-Ray Bursts}

\author[0000-0002-0173-6453]{N. Fraija}
\affiliation{Instituto de Astronom\' ia, Universidad Nacional Aut\'onoma de M\'exico,\\ Circuito Exterior, C.U., A. Postal 70-264, 04510 M\'exico City, M\'exico}

\author[0000-0003-4442-8546]{M. G. Dainotti}
\affiliation{National Astronomical Observatory of Japan, 2-21-1 Osawa, Mitaka, Tokyo 181-8588, Japan}
\affiliation{Space Science Institute, Boulder, CO, USA}
\affiliation{The Graduate University for Advanced Studies, SOKENDAI, Shonankokusaimura, Hayama, Miura District, Kanagawa 240-0193, Japan}
\footnote{First and second authors contributed equally to this paper}

\author[0000-0002-4764-1791]{Sahil Ugale}
\affiliation{Department of Physics, Mithibai College, Mumbai 400056, India}

\author{Debarpita Jyoti}
\affiliation{Department of Physics, Indian Institute of Technology Kharagpur, Kharagpur, West Bengal 721302, India}

\author[0000-0002-3222-9059]{Donald C. Warren}
\affiliation{RIKEN Interdisciplinary Theoretical and Mathematical Sciences Program (iTHEMS), Wak\={o}, Saitama, 351-0198 Japan}

\begin{abstract}
The Fermi Large Area Telescope (Fermi-LAT) Collaboration reported the Second Gamma-ray Burst Catalog (2FLGC), which comprises a subset of 29 bursts with photon energies above 10 GeV. Although the standard synchrotron forward-shock model has successfully explained the Gamma-ray burst (GRB) afterglow observations, energetic photons higher than 10 GeV from these transient events can hardly be described in this scenario. We present the closure relations (CRs) of synchrotron self-Compton (SSC) afterglow model in the adiabatic and radiative scenario and when the central engine injects continuous energy into the blastwave to study the evolution of the spectral and temporal indexes of those bursts reported in 2FLGC. We consider the SSC afterglow model evolving in stellar-wind and interstellar medium, and the CRs as a function of the radiative parameter, the energy injection index, and the electron spectral index for $1<p<2$ and $ 2\leq p$.  We select all GRBs that have been modeled with both a simple or a broken power law in the 2FLGC. We found that the CRs of the SSC model can satisfy a significant fraction of burst that cannot be interpreted in the synchrotron scenario, even though those that require an intermediate density profile (e.g., GRB 130427A) or an atypical fraction of total energy given to amplify the magnetic field ($\varepsilon_B$). The value of this parameter in the SSC model ranges ($\varepsilon_B\approx 10^{-5} - 10^{-4}$) when the cooling spectral break corresponds to the Fermi-LAT band for typical values of GRB afterglow. The analysis shows that ISM is preferred for the scenario without energy injection and the stellar wind medium for an energy injection scenario.
\end{abstract}

\section{Introduction}
Gamma-ray bursts (GRBs) are the most powerful transients in the Universe. These events lasting for milliseconds up to a pair of hours are classified in accordance with the duration of the main episode \citep{1993ApJ...413L.101K, 1999PhR...314..575P, 2015PhR...561....1K}, typically observed in the keV-MeV energy range and with spectra described by the empirical Band function \citep{1993ApJ...413..281B}. Following the prompt gamma-ray emission, a late long-lasting episode called ``afterglow" is detected from radio to GeV bands and interpreted when a relativistic outflow ejected by the central engine encounters the external environment and transfers a large part of its energy to the circumburst medium \citep{1997ApJ...476..232M, 1999PhR...314..575P}.   The shock-accelerated electrons in it are cooled down mainly by synchrotron radiation emitting photons from radio to gamma-rays \citep{1998ApJ...497L..17S,2009MNRAS.400L..75K, 2010MNRAS.409..226K,2013ApJ...763...71A, 2016ApJ...818..190F} and by synchrotron self-Compton (SSC) process up-scattering synchrotron photon at very-high energies \citep[VHE $\geq$ 10 GeV;][]{2019ApJ...885...29F, 2019ApJ...883..162F,2021ApJ...918...12F, 2017ApJ...848...94F,2019arXiv191109862Z}.

\cite{Ajello_2019} presented the Second Gamma-ray Burst Catalog (2FLGC), covering the first 10 years of operations (from 2008 to 2018 August 4). The catalog comprises 169 GRBs with high-energy emission higher than $\ge100$~ MeV, including a sample of GRBs with temporarily-extended emission and photon energies larger than a few GeV. The temporally extended emission was described in 2FLGC with a simple power-law (PL) and/or a broken power-law (BPL) function. Those bursts described with a BPL function showed a temporal break in the light curve around several hundreds of seconds. Synchrotron radiation from external forward shocks with the evolution of the temporal and spectral indexes is usually required to model the temporarily-extended emission, although not all the LAT light curves satisfy the relationship between these indices called  ``closure relations" (CRs) that are expected in case the forward shock dominates the high-energy emission \citep{1998ApJ...497L..17S,2009MNRAS.400L..75K,2010MNRAS.409..226K}.  Recently,  the temporal break in the temporarily-extended emission of GRB 160509A was interpreted as the passage of the synchrotron cooling break ($\nu^{\rm syn}_{\rm c}$) through the Fermi-LAT band \citep[$h\nu_{\rm LAT}=100\,{\rm MeV}$;][]{2020ApJ...905..112F}.\\

\cite{2019ApJ...883..134T} considered temporal ($\alpha_{\rm LAT}$) and spectral ($\beta_{\rm LAT}$) indexes and did a systematic analysis of the CRs in a sample of 59 selected LAT-detected bursts.  They found that although the standard synchrotron forward-shock emission describes the spectral and temporal indexes in most cases, there is still a considerable fraction of bursts that cannot be described with this model. They also found that several GRBs satisfy the CRs of the slow-cooling regime ($\nu^{\rm syn}_{\rm m}<\nu_{\rm LAT}<\nu^{\rm syn}_{\rm c}$) when the magnetic microphysical parameter has an atypical small value of $\epsilon_B<10^{-7}$. The term $\nu^{\rm syn}_{\rm m}$ corresponds to the characteristic spectral break of the synchrotron model. 

In addition, they found that long GRBs fulfilled the CRs derived in the constant-density medium. On the other hand,  \cite{2010MNRAS.403..926G} studied the high-energy emission of 11 Fermi LAT-detected bursts until 2009 October. They found that the LAT observations evolved as $\sim t^{-\frac{10}{7}}$ instead of $t^{-1}$ predicted by the adiabatic synchrotron forward-shock model, and therefore concluded that the relativistic forward shock was in the fully radiative regime for a hard spectral index $p\sim 2$.   \cite{2021ApJS..255...13D} explored in 2FLGC the presence of plateau phase naturally explained as continuous energy injection from the central engine into the blastwave \citep{2005ApJ...635L.133B,  2005ApJ...630L.113K, 2006Sci...311.1127D, 2006ApJ...636L..29P, 2006MNRAS.370L..61P, 2005Sci...309.1833B, 2007ApJ...671.1903C, 2017MNRAS.464.4399D, 2019ApJ...872..118B}. Authors found that three of them exhibited this phase, where the most favorable scenario was a slow-cooling regime in ISM.\\
%%%%
We derive the CRs of synchrotron self-Compton (SSC) afterglow model in the adiabatic and radiative scenario and when the central engine continuously injects energy into the blastwave to study its relation with bursts reported in 2FLGC. We consider the SSC afterglow model evolving in stellar-wind and constant-density medium, and the CRs as function of the radiative parameter $\epsilon$, the energy injection index $q$, and the electron spectral index for $1<p<2$ and $ 2\leq p$ (for the exact definition of $\epsilon$ and $q$, see the beginning of sections \ref{sec_211} and \ref{sec_22}, respectively).  We select all GRBs modeled with a SPL or a BPL in the 2FLGC. Out of 86 GRBs, the analysis shows the number and percentage of GRBs satisfying each relationship out of the CRs with/without energy injection.   The paper is arranged as follows. Section \S\ref{sec2} presents the  CRs of SSC afterglow models evolving in ISM and stellar-wind medium. In Section \S\ref{sec3}, we show Fermi-LAT data and the methodology. Section \S\ref{sec4} shows the analysis and discussion and finally,  in Section \S\ref{sec5}, we summarize.  

\section{Closure relations of SSC afterglow scenario}
\label{sec2}
The temporally extended afterglow emission is generated when a relativistic GRB outflow decelerates and drives a forward shock into the circumstellar medium. The outflow transfers a large amount of its energy to this surrounding external medium during the deceleration phase.  A large fraction of energy density is transferred during the shock to accelerate electrons ($\epsilon_e$) and amplify the magnetic field ($\epsilon_B$).   We consider that accelerated electrons are cooled down in a stratified external environment with a density profile described by $n(r) \propto r^{\rm -k}$. The density-profile indexes ${\rm k=0}$ 
and ${\rm k=2}$ correspond to ISM \cite[i.e, see][]{1998ApJ...497L..17S} and the stellar-wind environment
\cite[i.e, see][]{2000ApJ...536..195C}, respectively.  As follows, we estimate the CRs of SSC afterglow in i) adiabatic and radiative regime and ii) when the central engine continuously injects energy into the blastwave.

 %$n(r) = \frac{\dot{M}_{\rm W}}{4\pi v_{\rm W}}\, r^{\rm -k}$, where $\dot{M}_{\rm W}$ is the mass-loss rate and $v_{\rm W}$ is the wind velocity

\subsection{Adiabatic and radiative regime}

In the standard GRB afterglow scenario, the relativistic forward shock is usually considered to be fully adiabatic, although it can be partially or fully radiative \citep{1998MNRAS.298...87D, 1998ApJ...497L..17S, 2000ApJ...532..281B, 2010MNRAS.403..926G}. The shock-accelerated electrons are in the fast-cooling regime when the dynamical timescale is larger than the cooling timescale. In this case, the afterglow phase should be in the radiative regime instead of the adiabatic one \citep{2000ApJ...532..281B,2002MNRAS.330..955L,  2019ApJ...886..106P}. The deceleration of the GRB fireball by the external medium is faster when the fireball is radiative than adiabatic. Therefore,  the temporal afterglow evolution of synchrotron light curves and the shock's energetics are modified \citep{2000ApJ...532..281B, 2005ApJ...619..968W, 2000ApJ...529..151M}.

\subsubsection{Interstellar medium (${\rm k=0}$)}\label{sec_211}

Once the outflow begins to be decelerated at a significant distance from  the  progenitor by ISM, the evolution of the bulk Lorentz factor in the adiabatic and radiative regime  becomes $\Gamma=1.6\times 10^2 \left(\frac{1+z}{2}\right)^{\frac{3}{8-\epsilon}}\, n^{-\frac{1}{8-\epsilon}}E_{53}^{\frac{1}{8-\epsilon}}\Gamma_{0,2.7}^{-\frac{\epsilon}{8-\epsilon}}t_{2}^{-\frac{3}{8-\epsilon}}$, where $E$ is the equivalent kinetic energy, and $\Gamma_0$ is the initial Lorentz factor.   The parameter $\epsilon$ gives the account the hydrodynamic evolution of the forward shock in the fully adiabatic ($\epsilon=0$), fully radiative ($\epsilon=1$) or partially radiative or adiabatic ($0 <\epsilon < 1$) regimes.   The Lorentz factors, (i) of the lowest-energy electrons, (ii) above which electrons cool efficiently and, (iii) of the highest-energy electrons are

{\small
\begin{eqnarray}\nonumber
\label{ele_Lorent_ism}
\gamma_{\rm m}= \cases{
1.3\times 10^3\,\left(\frac{2-p}{p-1}\right)^{\frac{1}{p-1}}\left(\frac{1+z}{2}\right)^{\frac{3(4-p)}{2(8-\epsilon)(p-1)}}\, \epsilon^{\frac{1}{p-1}}_{e,-1}  \epsilon^{\frac{-(p-2)}{4(p-1)}}_{B,-2}  n^{\frac{8-6p+p\epsilon-2\epsilon}{4(p-1)(8-\epsilon)}} E_{53}^{\frac{4-p}{2(p-1)(8-\epsilon)}}\, \Gamma^{-\frac{\epsilon(4-p)}{2(8-\epsilon)(p-1)}}_{0,2.7}   t_{2}^{-\frac{3(4-p)}{2(8-\epsilon)(p-1)}}  \hspace{0.2cm} {\rm for} \hspace{0.2cm} { 1<p<2 }\cr
4.9\times 10^3\,\left(\frac{p-2}{p-1}\right) \left(\frac{1+z}{2}\right)^{\frac{3}{8-\epsilon}}\, \epsilon_{e,-1} n^{-\frac{1}{8-\epsilon}}\, E_{53}^{\frac{1}{8-\epsilon}}\, \Gamma^{-\frac{\epsilon}{8-\epsilon}}_{0,2.7}   t_{2}^{-\frac{3}{8-\epsilon}} \hspace{6.25cm} {\rm for} \hspace{0.2cm} {p > 2 }\cr
}
\end{eqnarray}
}
{\small
\bary\label{nus_ism}
\gamma_{\rm c}&=& 1.7\times 10^{2}\,(1+Y)^{-1} \left(\frac{1+z}{2}\right)^{-\frac{1+\epsilon}{8-\epsilon}}\, \epsilon^{-1}_{B,-2} n^{\frac{\epsilon-5}{8-\epsilon}}\, E_{53}^{-\frac{3}{8-\epsilon}}\, \Gamma^{\frac{3\epsilon}{8-\epsilon}}_{0,2.7}   t_{2}^{\frac{1+\epsilon}{8-\epsilon}}\,\cr
\gamma_{\rm max}&=& 1.9\times 10^{7}\, \left(\frac{1+z}{2}\right)^{-\frac{3}{2(8-\epsilon)}}\, \epsilon^{-\frac14}_{B,-2} n^{\frac{\epsilon-6}{4(8-\epsilon)}}\, E_{53}^{-\frac{1}{2(8-\epsilon)}}\, \Gamma^{\frac{\epsilon}{2(8-\epsilon)}}_{0,2.7}   t_{2}^{\frac{3}{8-\epsilon}}\,,\hspace{6.cm}
\eary
}

%depends on the maximum synchrotron flux, the optical depth $\tau\propto A_{\rm W}\,R^{-1}$ and the factor $k= 4\tilde{g}^{-1}$ for $1<p<2$ or $ 4 g^{-1}$ for $2\leq p$. The term $R$ is the forward-shock radius.  

respectively, where  $Y$ is the Compton parameter and $t$ is the observer time.  Hereafter, we adopt the convention $Q_{\rm x}=Q/10^{\rm x}$  in cgs units and use the values of p=1.9 and 2.2 for $1<p<2$ and $p>2$, respectively.   The sychrotron spectral breaks evolve as  $\nu^{\rm syn}_{\rm m}\propto t^{-\frac{3(p+2)}{(8-\epsilon)(p-1)}}$ for $1<p<2$, $\nu^{\rm syn}_{\rm m}\propto t^{-\frac{12}{8-\epsilon}}$ for $p> 2$ and $\nu^{\rm syn}_{\rm c}\propto t^{\frac{2(\epsilon-2)}{8-\epsilon}}$, and the maximum synchrotron flux as $F^{\rm syn}_{\rm max}\propto t^{-\frac{3\epsilon}{8-\epsilon}}$ \citep{2000ApJ...532..281B, 2005ApJ...619..968W}.\\  

The SSC process takes place when the same electron population that radiates synchrotron photons up-scatters them up to higher energies as $h\nu^{\rm ssc}_{\rm i}\simeq \gamma^2_{\rm i} h\nu^{\rm syn}_{\rm i}$, with ${\rm i=m, c}$ \citep[e.g., see][]{2001ApJ...548..787S}. The maximum flux that the SSC process is estimated as $F^{\rm ssc}_{\rm max}\sim \sigma_T n\,r\, F^{\rm syn}_{\rm max}$.\footnote{$\sigma_T$ is the Thompson cross section and $r$ is the forward-shock radius.} Given the electron Lorentz factors (Eq. \ref{ele_Lorent_ism}), the spectral breaks and the maximum flux of synchrotron process, the spectral breaks and the maximum flux in the SSC scenario and in the observer frame for $1<p<2$ and $p > 2$ are

{\small
\begin{eqnarray}\nonumber
\label{break_ssc_hom}
h\nu^{\rm ssc}_{\rm m} = \cases{ 
1.4\times 10^7\,{\rm eV}\,\left(\frac{2-p}{p-1}\right)^{\frac{4}{p-1}} \left(\frac{1+z}{2}\right)^{\frac{26+p(\epsilon-8)-\epsilon}{(8-\epsilon)(p-1)}}\, \epsilon^{\frac{4}{p-1}}_{e,-1}  \epsilon^{\frac{3-p}{2(p-1)}}_{B,-2}  n^{\frac{p(\epsilon-8)-3(\epsilon-4)}{2(p-1)(8-\epsilon)}}  E_{53}^{\frac{6}{(p-1)(8-\epsilon)}}\, \Gamma^{-\frac{6\epsilon}{(8-\epsilon)(p-1)}}_{0,2.7}   t_{2}^{-\frac{18}{(8-\epsilon)(p-1)}}  \hspace{0.2cm} {\rm for} \hspace{0.2cm} { 1<p<2 }\cr
3.4\times 10^9\,{\rm eV}\,\left(\frac{p-2}{p-1}\right)^4 \left(\frac{1+z}{2}\right)^{\frac{10+\epsilon}{8-\epsilon}}\, \epsilon^4_{e,-1} \epsilon^{\frac12}_{B,-2} n^{-\frac{4+\epsilon}{2(8-\epsilon)}}\, E_{53}^{\frac{6}{8-\epsilon}}\, \Gamma^{-\frac{6\epsilon}{8-\epsilon}}_{0,2.7}   t_{2}^{-\frac{18}{8-\epsilon}} \hspace{4.85cm} {\rm for} \hspace{0.2cm} {p > 2 }\cr
}
\end{eqnarray}
}

{\small
\bary \label{ssc_br_hom_v1}
h\nu^{\rm ssc}_{\rm c}&=& 1.4\times 10^3\,{\rm eV} \left(\frac{1+z}{2}\right)^{-4} (1+z)^{-\frac{3(2+\epsilon)}{8-\epsilon}}\, \epsilon^{-\frac72}_{B,-2} n^{\frac{7\epsilon-36}{2(8-\epsilon)}}\, E_{53}^{-\frac{10}{8-\epsilon}}\, \Gamma^{\frac{10\epsilon}{8-\epsilon}}_{0,2.7} \,t_{2}^{\frac{2(2\epsilon-1)}{8-\epsilon}}\cr
F^{\rm ssc}_{\rm max}&=& 7.1\times 10^{-5}\,{\rm mJy} \left(\frac{1+z}{2}\right)^{\frac{2(\epsilon+7)}{8-\epsilon}}\,  \epsilon^\frac12_{B,-2} D_{\rm z,27.3}^{-2}\,n^{\frac{5(4-\epsilon)}{2(8-\epsilon)}}\, E_{53}^{\frac{10}{8-\epsilon}}\, \Gamma^{-\frac{10\epsilon}{8-\epsilon}}_{0,2.7}   t_{2}^{\frac{2(1-2\epsilon)}{8-\epsilon}}\,,
\eary
}

respectively. It is worth noting that the spectral breaks and the maximum flux derived in \cite{2019ApJ...883..162F} are recovered for $\epsilon=0$. Given the evolution of the spectral breaks and the maximum flux (Eqs. \ref{ssc_br_hom_v1}), we estimate the CRs of the SSC model in ISM as listed in Table \ref{ClosureRelations}.

\subsubsection{Stellar-wind medium (${\rm k=2}$)}
 Relativistic electrons are accelerated in the forward shock and cooled efficiently by synchrotron and SSC processes. In general,  at a considerable distance from the progenitor, the evolution in the adiabatic and radiative regime of the bulk Lorentz factor is $\Gamma=86.7 \left(\frac{1+z}{2}\right)^{\frac{1}{4-\epsilon}}\, A_{\rm W,-1}^{-\frac{1}{4-\epsilon}}E_{53}^{\frac{1}{4-\epsilon}}\Gamma_{0,2.7}^{-\frac{\epsilon}{4-\epsilon}}t_{2}^{-\frac{1}{4-\epsilon}}$, where the term $A_{\rm W}$ corresponds to the density parameter \citep{2000ApJ...536..195C}.  The Lorentz factors, (i) of the lowest-energy electrons, (ii) above which electrons cool efficiently and, (iii) of the highest-energy electrons are

{\small
\begin{eqnarray}\nonumber
\label{ele_win}
\gamma_{\rm m}= \cases{ 
6.5\times 10^2\,\left(\frac{2-p}{p-1}\right)^{\frac{1}{p-1}}\left(\frac{1+z}{2}\right)^{\frac{8+p(\epsilon-3)-2\epsilon}{2(4-\epsilon)(p-1)}}\, \epsilon^{\frac{1}{p-1}}_{e,-1}  \epsilon^{\frac{-(p-2)}{4(p-1)}}_{B,-2}  A_{\rm W,-1}^{\frac{8+p(\epsilon-6)-2\epsilon }{4(p-1)(4-\epsilon)}}E_{53}^{\frac{p}{2(p-1)(4-\epsilon)}}\, \Gamma^{-\frac{p\epsilon}{2(4-\epsilon)(p-1)}}_{0,2.7} t_{2}^{\frac{2\epsilon+ p ( 3 - \epsilon) - 8}{2(4-\epsilon)(p-1)}} \hspace{0.7cm} {\rm for} \hspace{0.2cm} { 1<p<2 }\cr
%&& \hspace{0.1cm}\,     \hspace{0.8cm} \cr
2.7\times 10^3\,\left(\frac{p-2}{p-1}\right) \left(\frac{1+z}{2}\right)^{\frac{1}{4-\epsilon}}\, \epsilon_{e,-1} \,  A_{\rm W,-1}^{-\frac{1}{4-\epsilon}}\, E_{53}^{\frac{1}{4-\epsilon}}\, \Gamma^{-\frac{\epsilon}{4-\epsilon}}_{0,2.7}   t_{2}^{-\frac{1}{4-\epsilon}} \hspace{6.75cm} {\rm for} \hspace{0.2cm} {p> 2 }
}
\end{eqnarray}
}
{\small
\bary\label{nus_wind}
\gamma_{\rm c}&=& 72.9\,(1+Y)^{-1} \left(\frac{1+z}{2}\right)^{\frac{\epsilon-3}{4-\epsilon}}\,\epsilon^{-1}_{B,-2}    A_{\rm W,-1}^{\frac{\epsilon-5}{4-\epsilon}}E_{53}^{\frac{1}{4-\epsilon}}\,\Gamma^{-\frac{\epsilon}{4-\epsilon}}_{0,2.7}   t_{2}^{\frac{3-\epsilon}{4-\epsilon}},\cr
\gamma_{\rm max}&=& 1.3\times 10^7\, \left(\frac{1+z}{2}\right)^{\frac{\epsilon-3}{2(4-\epsilon)}}\,\epsilon^{-\frac14}_{B,-2}    A_{\rm W,-1}^{\frac{\epsilon-6}{4(4-\epsilon)}}E_{53}^{\frac{1}{2(4-\epsilon)}}\,\Gamma^{-\frac{\epsilon}{2(4-\epsilon)}}_{0,2.7}   t_{2}^{\frac{3-\epsilon}{2(4-\epsilon)}},\hspace{7cm}
\eary
}

%where  $\epsilon_{\rm e}$ and $\epsilon_{\rm B}$ are the fraction of  the total energy going to electrons and magnetic field, respectively.
%where $\tilde{g}=\frac{2-p}{p-1}$, and $g=\frac{p-2}{p-1}$ for $2<p$ and ${\rm ln^{-1}}\left( \frac{\gamma_M}{\gamma_m} \right)$ for p=2, with $\gamma_{\rm M}$ the maximum electron Lorentz factor
respectively, with $t$ the observer time.  The synchrotron spectral breaks in the stellar-wind afterglow model evolve as  $\nu^{\rm syn}_{\rm m}\propto t^{\frac{\epsilon-p-4}{(4-\epsilon)(p-1)}}$ for $1<p<2$, $\nu^{\rm syn}_{\rm m}\propto t^{\frac{\epsilon-6}{4-\epsilon}}$ for $p> 2$ and $\nu^{\rm syn}_{\rm c}\propto t^{\frac{2-\epsilon}{4-\epsilon}}$, and the maximum synchrotron flux as $F^{\rm syn}_{\rm max}\propto t^{-\frac{2}{4-\epsilon}}$.

Given the electron Lorentz factors (Eq. \ref{nus_wind}), the spectral breaks and the maximum flux of synchrotron process, the spectral breaks and the maximum flux in the SSC scenario and in the observer frame for $1<p<2$ and $p>2$ are

{\small
\begin{eqnarray}\nonumber
\label{break_ssc_win}
h\nu^{\rm ssc}_{\rm m}= \cases{ 
1.2\times 10^6\,{\rm eV}\, \left(\frac{2-p}{p-1}\right)^{\frac{4}{p-1}}\left(\frac{1+z}{2}\right)^{\frac{2[8+p(\epsilon-3)  - 2 \epsilon]}{(4-\epsilon)(p-1)}} \epsilon^{\frac{4}{p-1}}_{e,-1}  \epsilon^{\frac{3-p}{2(p-1)}}_{B,-2} A_{\rm W,-1}^{\frac{p(\epsilon - 8) +  3(4 - \epsilon) }{2(p-1)(4-\epsilon)}}\, E_{53}^{\frac{2p}{(p-1)(4-\epsilon)}} \Gamma^{-\frac{2p\epsilon}{(4-\epsilon)(p-1)}}_{0,2.7} t_{2}^{\frac{3(\epsilon - 4) -p (\epsilon - 2)}{(4-\epsilon)(p-1)}} \hspace{0.1cm} {\rm for} \hspace{0.2cm} { 1<p<2 }\cr
3.2\times10^8\,{\rm eV}\,\left(\frac{p-2}{p-1}\right)^{4}\left(\frac{1+z}{2}\right)^{\frac{4}{4-\epsilon}}\, \epsilon^{\frac12}_{B,-2} \epsilon^{4}_{e,-1}\,  A_{\rm W,-1}^{-\frac{4+\epsilon}{2(4-\epsilon)}}\, E_{53}^{\frac{4}{4-\epsilon}}\, \Gamma^{-\frac{4\epsilon}{4-\epsilon}}_{0,2.7} t_{2}^{\frac{\epsilon-8}{4-\epsilon}} \hspace{5.4cm} {\rm for} \hspace{0.2cm} {p > 2}\cr
}
\end{eqnarray}
}
{\small
\bary \label{ssc_br_win_eps}
h\nu^{\rm ssc}_{\rm c}&=& 55.1\,{\rm eV} (1+Y)^{-4} \left(\frac{1+z}{2}\right)^{\frac{4(\epsilon-3)}{4-\epsilon}}\, \epsilon^{-\frac72}_{B,-2}   A_{\rm W,-1}^{\frac{7\epsilon-36}{2(4-\epsilon)}}\, E_{53}^{\frac{4}{4-\epsilon}}\, \Gamma^{-\frac{4\epsilon}{4-\epsilon}}_{0,2.7}   t_{2}^{\frac{8-3\epsilon}{4-\epsilon}}\cr
F^{\rm ssc}_{\rm max}&=& 1.2\times 10^{-4}\,{\rm mJy} \left(\frac{1+z}{2}\right)^3\, \epsilon^{\frac12}_{B,-2}\, D_{\rm z,27.3}^{-2} A_{\rm W,-1}^{\frac{5}{2}}\,  t_{2}^{-1}\,.\hspace{9cm}
\eary
}

Given the evolution of the spectral breaks and the maximum flux (Eqs. \ref{ssc_br_win_eps}), we estimate the CRs of SSC model in stellar-wind environment as listed in Table \ref{ClosureRelations}. It is worth noting that the spectral breaks and the maximum flux derived in \citep{2015ApJ...804..105F, 2019ApJ...883..162F} are recovered for $\epsilon=0$

%Given the evolution of the spectral breaks and the maximum flux (Eqs. \ref{ssc_br_win}), it is possible to write the SSC LCs as reported in Table \ref{ClosureRelations}.

\begin{table}[h!]
\centering \renewcommand{\arraystretch}{1.85}\addtolength{\tabcolsep}{1pt}
\caption{CRs of the SSC afterglow model in ISM and stellar-wind for an adiabatic and radiative scenario}
\label{ClosureRelations}
\begin{tabular}{ccccccc}
\hline
 &  & $\beta$ & $\alpha(p)$ & $\alpha(p)$ & $\alpha(\beta)$ & $\alpha(\beta)$ \\
 &  & & $(1<p<2)$ & $(p>2)$ & $(1<p<2)$ & $(p > 2)$ \\
\hline
 &  &  & ISM, Slow cooling &  &  &  \\
\hline
1 & $\nu^{\rm ssc}_m<\nu<\nu^{\rm ssc}_c$ & $\frac{p-1}{2}$ & $\frac{7+4\epsilon}{8-\epsilon}$ & $-\frac{11-4\epsilon-9p}{8-\epsilon}$ & ----- & $\frac{2(2\epsilon+9\beta-1)}{8-\epsilon}$ \\
2 & $\nu^{\rm ssc}_c<\nu$ & $\frac{p}{2}$ & $\frac{2(4+\epsilon)}{8-\epsilon}$ & $-\frac{10 - 2\epsilon-9p}{8-\epsilon}$ & ----- & $\frac{2(\epsilon+9\beta-5)}{8-\epsilon}$ \\
\hline
 &  &  & ISM, Fast cooling &  &  &  \\
\hline
3 & $\nu^{\rm ssc}_c<\nu<\nu^{\rm ssc}_m$ & $\frac{1}{2}$ & $-\frac{1-2\epsilon}{8-\epsilon}$ & $-\frac{1-2\epsilon}{8-\epsilon}$ & $\frac{2(2\epsilon-1)\beta}{8-\epsilon}$ & $\frac{2(2\epsilon-1)\beta}{8-\epsilon}$ \\
4 & $\nu^{\rm ssc}_m<\nu$ & $\frac{p}{2}$ & $\frac{2(4+\epsilon)}{8-\epsilon}$ & $-\frac{10 - 2\epsilon-9p}{8-\epsilon}$ & ----- & $\frac{2(\epsilon+9\beta-5)}{8-\epsilon}$ \\
\hline
&  &  & Wind, Slow cooling &  &  &  \\
\hline
5 & $\nu^{\rm ssc}_m<\nu<\nu^{\rm ssc}_c$ & $\frac{p-1}{2}$ &$-\frac{2p+5\epsilon - 20 - p\epsilon }{2(4-\epsilon)}$ & $-\frac{p(\epsilon - 8) + \epsilon}{2(4-\epsilon)}$ & $\frac{9-2\epsilon+\beta(\epsilon-2)}{4-\epsilon}$ & $\frac{4-\epsilon+\beta(8-\epsilon)}{4-\epsilon}$ \\
6 & $\nu^{\rm ssc}_c<\nu$ & $\frac{p}{2}$ & $-\frac{2(\epsilon-6)-p(\epsilon-2)}{2(4-\epsilon)}$ & $-\frac{8+p(\epsilon-8)-2\epsilon}{2(4-\epsilon)}$ & $\frac{6-\epsilon+\beta(\epsilon-2)}{4-\epsilon}$ & $\frac{\epsilon-4+\beta(8-\epsilon)}{4-\epsilon}$ \\
\hline
 &  &  & Wind, Fast cooling &  &  &  \\
\hline
7 & $\nu^{\rm ssc}_c<\nu<\nu^{\rm ssc}_m$ & $\frac{1}{2}$ & $\frac{\epsilon}{2(4-\epsilon)}$ & $\frac{\epsilon}{2(4-\epsilon)}$ & $\frac{\epsilon\beta}{4-\epsilon}$ & $\frac{\epsilon\beta}{4-\epsilon}$ \\
8 & $\nu^{\rm ssc}_m<\nu$ & $\frac{p}{2}$ & $-\frac{2(\epsilon-6)-p(\epsilon-2)}{2(4-\epsilon)}$ & $-\frac{8+p(\epsilon-8)-2\epsilon}{2(4-\epsilon)}$ & $\frac{6-\epsilon+\beta(\epsilon-2)}{4-\epsilon}$ & $\frac{\epsilon-4+\beta(8-\epsilon)}{4-\epsilon}$ \\
\hline
\end{tabular}
%Note: 
\end{table}

\subsection{Energy injection scenario}\label{sec_22}

Continuous energy injection by the central engine on the GRB afterglow can produce refreshed shocks. A continuous luminosity of the central engine can be described by  $L_{\rm inj}(t)\propto L_0\, t^{\rm -q}$  where $q$ is the energy injection index and $L_{\rm inj}$ is the luminosity injected into the blastwave \citep[e.g.][]{2006ApJ...642..354Z}. The equivalent kinetic energy can be estimated as $E =\int L_{\rm inj}\, dt\propto L_0 t^{-q+1}$.

\subsubsection{Interstellar medium (${\rm k=0}$)}

The evolution of the bulk Lorentz factor in ISM becomes  {\small $\Gamma=1.7\times 10^2\left(\frac{1+z}{2}\right)^{\frac{3}{8}}\,  n^{-\frac{1}{8}}E_{53}^{\frac{1}{8}} t_{2}^{-\frac{q+2}{8}}$}. The Lorentz factors, (i) of the lowest-energy electrons, (ii) above which electrons cool efficiently and, (iii) of the highest-energy electrons are

{\small
\begin{eqnarray}\label{gam_m_i}  
\gamma_{\rm m}= \cases{
1.3\times 10^3\,\left(\frac{2-p}{p-1}\right)^\frac{1}{p-1} \left(\frac{1+z}{2}\right)^{\frac{3(4-p)}{16(p-1)}}\, \epsilon^{\frac{1}{p-1}}_{e,-1}  \epsilon^{\frac{-(p-2)}{4(p-1)}}_{B,-2}  n^{\frac{4-3p}{16(p-1)}} E_{53}^{\frac{4-p}{16(p-1)}}\,t_{2}^{\frac{(p-4)(q+2)}{16(p-1)}}  \hspace{0.2cm} {\rm for} \hspace{0.2cm} { 1<p<2 }\,\,\,\,\,\,\,\cr
5.1\times 10^3\,\left(\frac{p-2}{p-1}\right)\left(\frac{1+z}{2}\right)^{\frac{3}{8}}\, \epsilon_{e,-1} n^{-\frac{1}{8}}\, E_{53}^{\frac{1}{8}}\,  t_{2}^{-\frac{2+q}{8}} \hspace{4.4cm} {\rm for} \hspace{0.2cm} {p> 2 }\cr
}
\end{eqnarray}
}

{\small
\begin{eqnarray}\label{nu_c_i}
\gamma_{\rm c}&=& 1.5\times 10^2\,(1+Y)^{-1} \left(\frac{1+z}{2}\right)^{-\frac{1}{8}}\, \epsilon^{-1}_{B,-2} n^{-\frac{5}{8}}\, E_{53}^{-\frac{3}{8}}\,  t_{2}^{\frac{3q-2}{8}}\cr
\gamma_{\rm max}&=& 1.8\times 10^7\,{\rm mJy} \left(\frac{1+z}{2}\right)^{-\frac{3}{16}}\,  \varepsilon^{-\frac14}_{B,-2}\,n^{-\frac{3}{16}}\, E^{-\frac{1}{16}}_{53}\,  t_{4.7}^{\frac{q+2}{16}}\,,\hspace{7cm}
\end{eqnarray}
}
respectively. The synchrotron spectral breaks evolve as $\nu^{\rm syn}_{\rm m}\propto t^{-\frac{(p+2)(q+2)}{8(p-1)}}$  for $1<p<2$, $\nu^{\rm syn}_{\rm m}\propto t^{-\frac{2+q}{2}}$ for $p> 2$ and $\nu^{\rm syn}_{\rm c}\propto t^{\frac{q-2}{2}}$, and maximum synchrotron flux as $F^{\rm syn}_{\rm max}\propto t^{1-q}$.   Given the electron Lorentz factors, the synchrotron spectral breaks and the maximum synchrotron flux, the spectral breaks and the maximum flux in the SSC scenario and in the observer frame are

{\small
\begin{eqnarray}\nonumber
\label{syn_esp_win}
h\nu^{\rm ssc}_{\rm m} = \cases{ 
1.8\times 10^{7}\,{\rm eV} \left(\frac{2-p}{p-1}\right)^{\frac{4}{p-1}} \left(\frac{1+z}{2}\right)^{\frac{13 - 4p}{4(p-1)}}\, \epsilon^{\frac{4}{p-1}}_{e,-1}  \epsilon^{\frac{3-p}{2(p-1)}}_{B,-2}  n^{\frac{3 - 2p }{4(p-1)}}  E^{\frac{3}{4(p-1)}}_{53}\,t_{2}^{-\frac{3(q+2)}{4(p-1)}}  \hspace{0.7cm} {\rm for} \hspace{0.2cm} { 1<p<2 }\cr
4.2\times 10^9\,{\rm eV}\left(\frac{p-2}{p-1}\right)^4\left(\frac{1+z}{2}\right)^{\frac{5}{4}}\, \epsilon^4_{e,-1} \epsilon^{\frac12}_{B,-2} n^{-\frac{1}{4}}\, E^{\frac{3}{4}}_{53}\,   t_{2}^{-\frac{3(2+q)}{4}} \hspace{2.9cm} {\rm for} \hspace{0.2cm} {p> 2 }\cr
}
\end{eqnarray}
}

{\small
\bary \label{ssc_br_hom_q}
h\nu^{\rm ssc}_{\rm c}&=& 9.8\times 10^2\,{\rm eV} (1+Y)^{-4} \left(\frac{1+z}{2}\right)^{-\frac{3}{4}}\, \epsilon^{-\frac72}_{B,-2} n^{-\frac{9}{4}}\, E_{53}^{-\frac{5}{4}}\, t_{2}^{\frac{5q-6}{4}}\cr
F^{\rm ssc}_{\rm max}&=& 1.1\times 10^{-4}\,{\rm mJy} \left(\frac{1+z}{2}\right)^{\frac{7}{4}}\,  \epsilon^\frac12_{B,-2} D_{\rm z,27.3}^{-2}\,n_{-1}^{\frac{5}{4}}\, E_{53}^{\frac{5}{4}}\, t_{2}^{\frac{6-5q}{4}}\,,
\eary
}

respectively.  Given the evolution of the spectral breaks and the maximum flux (Eq. \ref{ssc_br_hom_q}), we estimate the CRs of the SSC model in ISM as reported in Table \ref{ClosureRelations2}.

\subsubsection{Stellar-wind medium (${\rm k=2}$)}

The evolution of the bulk Lorentz factor during the deceleration phase in stellar-wind medium is {\small $\Gamma= 1.2\times 10^2\left(\frac{1+z}{2}\right)^{\frac{1}{4}}\,   A_{\rm W,-1}^{-\frac{1}{4}}E_{53}^{\frac{1}{4}}t_{2}^{-\frac{q}{4}}$}.    The Lorentz factors, (i) of the lowest-energy electrons, (ii) above which electrons cool efficiently and, (iii) of the highest-energy electrons are

{\small
\begin{eqnarray}\label{gam_m_w}
\gamma_{\rm m}=\cases{
8.8\times 10^2\,\left(\frac{2-p}{p-1}\right)^{\frac{1}{p-1}}\left(\frac{1+z}{2}\right)^{\frac{8-3p}{4(p-1)}}\,  \epsilon^{\frac{1}{p-1}}_{e,-1}  \epsilon^{\frac{-(p-2)}{4(p-1)}}_{B,-2}  A_{\rm W,-1}^{\frac{4-3p}{8(p-1)}}E_{53}^{\frac{p}{8(p-1)}}\, t_{2}^{-\frac{8 + p(q - 4)}{8(p-1)}} \hspace{0.3cm} {\rm for} \hspace{0.2cm} { 1<p<2 }\,\,\,\,\,\,\cr
3.5\times 10^3\, \left(\frac{p-2}{p-1}\right)\left(\frac{1+z}{2}\right)^{\frac{1}{4}}\, \,\epsilon_{e,-1} \,  A_{\rm W,-1}^{-\frac{1}{4}}\, E_{53}^{\frac{1}{4}}\, t_{2}^{-\frac{q}{4}} \hspace{4.1cm} {\rm for} \hspace{0.2cm} {p> 2}}
\end{eqnarray}}

{\small
\begin{eqnarray}\label{nu_c_w}
\gamma_{\rm c}&=&97.1 (1+Y)^{-1} \left(\frac{1+z}{2}\right)^{-\frac{3}{4}}\,\epsilon^{-1}_{B,-2}    A_{\rm W,-1}^{-\frac{5}{4}}E_{53}^{\frac{1}{4}}\, t_{2}^{\frac{4-q}{4}},\cr
\gamma_{\rm max}&=&1.5\times 10^7\left(\frac{1+z}{2}\right)^{-\frac{3}{8}}\,\epsilon^{-\frac{1}{4}}_{B,-2}    A_{\rm W,-1}^{-\frac{3}{8}}E^{\frac18}_{53}\, t_{2}^{\frac{4-q}{8}}\,,\hspace{7.2cm}
\end{eqnarray}}
respectively.   The synchrotron spectral breaks evolve as $\nu^{\rm syn}_{\rm m}\propto t^{-\frac{4+pq}{4(p-1)}}$  for $1<p<2$, $\nu^{\rm syn}_{\rm m}\propto t^{-\frac{2+q}{2}}$ for $p> 2$ and $\nu^{\rm syn}_{\rm c}\propto t^{-\frac{q-2}{2}}$, and the maximum spectral breaks $F^{\rm syn}_{\rm max}\propto t^{-\frac{q}{2}}$.\\

%Given the electron Lorentz factors, the synchrotron spectral breaks and the maximum synchrotron flux, the spectral breaks and the maximum flux in the SSC scenario are

The spectral breaks and maximum flux in the SSC scenario are determined by the electron Lorentz factors. In this case, the synchrotron spectral breaks, and maximum synchrotron flux in the observer frame are

{\small
\begin{eqnarray}\nonumber
\label{ssc_esp_win}
h\nu^{\rm ssc}_{\rm m}= \cases{ 
3.9\times 10^{6}\,{\rm eV} \left(\frac{2-p}{p-1}\right)^{\frac{4}{p-1}} \left(\frac{1+z}{2}\right)^{\frac{8-3p}{2(p-1)}} \epsilon^{\frac{4}{p-1}}_{e,-1}  \epsilon^{\frac{3-p}{2(p-1)}}_{B,-2}   A_{\rm W,-1}^{\frac{3-2p}{2(p-1)}}\, E_{53}^{\frac{p}{2(p-1)}}\, t_{2}^{-\frac{6+p(q-2)}{2(p-1)}} \hspace{0.4cm} {\rm for} \hspace{0.2cm} { 1<p<2 }\cr
1.0\times 10^9\,{\rm eV}\,\left(\frac{p-2}{p-1}\right)^4 \left(\frac{1+z}{2}\right)\, \epsilon^{\frac12}_{B,-2} \epsilon^{4}_{e,-1}\,   A_{\rm W,-1}^{-\frac{1}{2}}\, E_{53}\, t_{2}^{-(q+1)} \hspace{3cm} {\rm for} \hspace{0.1cm} {p> 2 }\cr
}
\end{eqnarray}
}

{\small
\bary \label{ssc_br_win}
h\nu^{\rm ssc}_{\rm c}&=& 1.7\times 10^2\,{\rm GeV} (1+Y)^{-4} \left(\frac{1+z}{2}\right)^{-3}\, \epsilon^{-\frac72}_{B,-2}  \,  A_{\rm W,-1}^{-\frac{9}{2}}\, E_{53}\, t_{2}^{3-q}\cr
F^{\rm ssc}_{\rm max}&=& 1.2\times 10^{-4}\,{\rm mJy} \left(\frac{1+z}{2}\right)^3\, \epsilon^{\frac12}_{B,-2} \, D_{\rm z,27.3}^{-2} A_{\rm W,-1}^{\frac{5}{2}}\,  t_{2}^{-1}\,.\hspace{6.5cm}
\eary
}

%Given the evolution of the spectral breaks and the maximum flux (Eq. \ref{ssc_esp_win}), we estimate the CRs of the SSC model in stellar-wind medium as reported in Table \ref{ClosureRelations2}.

We estimate the CRs of the SSC model in stellar-wind medium using the evolution of the spectral breaks and the maximum flux (Eq. \ref{ssc_esp_win}), as shown in Table \ref{ClosureRelations2}.

\begin{table}[h!]
\centering \renewcommand{\arraystretch}{1.85}\addtolength{\tabcolsep}{1pt}
\caption{CRs of the SSC afterglow model in ISM and stellar-wind for energy injection scenario}
\label{ClosureRelations2}
\begin{tabular}{ccccccc}
\hline
 &  & $\beta$ & $\alpha(p)$ & $\alpha(p)$ & $\alpha(\beta)$ & $\alpha(\beta)$ \\
 &  & & $(1<p<2)$ & $ (p>2)$ & $(1<p<2)$ & $ (p>2)$ \\
\hline
 &  &  & ISM, Slow cooling &  &  &  \\
\hline
1 & $\nu^{\rm ssc}_m<\nu<\nu^{\rm ssc}_c$ & $\frac{p-1}{2}$ & $-\frac{6-13q}{8}$ & $-\frac{18-7q-6p-3qp}{8}$ & ----- & $\frac{3\beta(q+2)+5q-6}{4}$ \\
2 & $\nu^{\rm ssc}_c<\nu$ & $\frac{p}{2}$ & $q$ & $-\frac{12-2q-6p-3pq}{8}$ & ----- & $\frac{3\beta(q+2)+q-6}{4}$ \\
\hline
 &  &  & ISM, Fast cooling &  &  &  \\
\hline
3 & $\nu^{\rm ssc}_c<\nu<\nu^{\rm ssc}_m$ & $\frac{1}{2}$ & $-\frac{6-5q}{8}$ & $-\frac{(6-5q)}{8}$ & $\frac{\beta(5q-6)}{4}$ & $\frac{\beta(5q-6)}{4}$ \\
4 & $\nu^{\rm ssc}_m<\nu$ & $\frac{p}{2}$ & $q$ & $-\frac{12-2q-6p-3pq}{8}$ & ----- & $\frac{3\beta(q+2)+q-6}{4}$ \\
\hline
&  &  & Wind, Slow cooling &  &  &  \\
\hline
5 & $\nu^{\rm ssc}_m<\nu<\nu^{\rm ssc}_c$ & $\frac{p-1}{2}$ & $-\frac{2p-pq-10}{4}$ & $-\frac{q-1-p-pq}{2}$ & $\frac{2\beta(q-2)+q+8}{4}$ & $\beta(q+1)+1$ \\
6 & $\nu^{\rm ssc}_c<\nu$ & $\frac{p}{2}$ & $\frac{p(q-2)+2(q+2)}{4}$ & $-\frac{2-p-pq}{2}$ & $\frac{\beta(q-2)+q+2}{2}$ & $\beta(q+1)-1$ \\
\hline
 &  &  & Wind, Fast cooling &  &  &  \\
\hline
7 & $\nu^{\rm ssc}_c<\nu<\nu^{\rm ssc}_m$ & $\frac{1}{2}$ & $-\frac{1-q}{2}$ & $-\frac{1-q}{2}$ & $\beta(q-1)$ & $\beta(q-1)$ \\
8 & $\nu^{\rm ssc}_m<\nu$ & $\frac{p}{2}$ & $\frac{p(q-2)+2(q+2)}{4}$ & $-\frac{2-p-pq}{2}$ & $\frac{\beta(q-2)+q+2}{2}$ & $\beta(q+1)-1$ \\
\hline

\end{tabular}
\end{table}

\section{The methodology}
\label{sec3}
We select GRBs from the 2FLGC \citep{Ajello_2019}: 86 with temporally extended emission with durations lasting from 31 s to 34366 s \citep{2009MNRAS.400L..75K, 2010MNRAS.409..226K, 2016ApJ...818..190F}. In 2FLGC, the temporal emission is fitted with a SPL function, and GRBs with four or more flux measurements (that are not simply upper limits) are additionally fitted with a BPL.

For 21 bursts for which it was possible to fit the PL and the BPL, we use the value of $\alpha_2$ from the BPL fit, as the BPL fit allows us to investigate the CR later.   For the analysis of the CRships, we follow our previous strategy presented in \cite{2021PASJ...73..970D, 2021ApJS..255...13D, 2020ApJ...903...18S}.  The errorbars quoted in the plots are in the 1 $\sigma$ range, and we consider the errorbars among the $\alpha$ and $\beta$ values to be dependent; thus, the figures in the result sections are shown with ellipses. 

Our sample consists of 86 total GRBs -- 65 fitted with a PL and 21 fitted with a BPL. When analyzing the CRs, we use the temporal and spectral indices according to the relationship $F_{\rm \nu} \propto t^{-\alpha} \nu^{-\beta}$. We use the temporal indexes $\alpha$ and $\alpha_2$ from the 2FLGC when considering the PL and BPL, respectively. The spectral index $\beta$ is taken as the photon index from the catalog minus one.   The distributions of $\alpha$ and $\beta$ are roughly Gaussian, with most values of $\alpha$ ranging between $0.25$ and $2.75$, and most values of $\beta$ ranging between $0.6$ and $1.8$.

We have derived and listed in Tables \ref{ClosureRelations} and \ref{ClosureRelations2} the CRs of SSC afterglow model in the adiabatic and radiative scenario, and when the central engine continuously injects energy into the blastwave.  We consider the SSC afterglow model evolving in ISM and stellar-wind medium, and the CRs as function of the radiative parameter $\epsilon$, the energy injection index $q$, and the electron spectral index for $1<p<2$ and $ 2\leq p$. 

%The CRs of SSC model in the radiative regime without energy injection and with energy injection evolving in ISM for $1<p<2$ with the cooling conditions of ${\rm \nu^{\rm ssc}_{\rm m}\leq \nu \leq  \nu^{\rm ssc}_{\rm c}}$, and ${\rm \{\nu^{\rm ssc}_{\rm m}, \nu^{\rm ssc}_{\rm c}\}<\nu}$ are not satisfied, and therefore these are not estimated.\\ 

Based on the CRs of SSC afterglow models, we here discuss the results in terms of the bursts reported in 2FLGC.

\begin{table}[h]
    \centering\renewcommand{\arraystretch}{1.2}
    \begin{tabular}{c c c c c c c}
    \toprule[1.2pt]
    \toprule[1.2pt]
     Type & Cooling & $\nu$ Range & CR: p $>$ 2 & CR: 1 $<$ p $<$ 2 & GRBs Satisfying Relation & Proportion Satisfying Relation \\
     \midrule
     ISM & slow & ${\rm \nu_m^{ssc} < \nu < \nu_{c}^{ssc}}$ & $\frac{2(2\epsilon+9\beta-1)}{8-\epsilon}$ & $-$ & 3 & 3.49\% \\
    \hline
    ISM & slow & ${\rm \nu_c^{ssc} < \nu}$ & $\frac{2(\epsilon+9\beta-5)}{8-\epsilon}$ & $-$ & 18 & 20.9\% \\
    \hline
    ISM & fast & ${\rm \nu_c^{ssc} < \nu < \nu_{m}^{ssc}}$ & $\frac{2(2\epsilon-1)\beta}{8-\epsilon}$ & $\frac{2(2\epsilon-1)\beta}{8-\epsilon}$ & 2 & 2.33\% \\
    \hline
    ISM & fast & ${\rm \nu_m^{ssc} < \nu}$ & $\frac{2(\epsilon+9\beta-5)}{8-\epsilon}$ & $-$ & 18 & 20.9\% \\
    \hline
    Wind & slow & ${\rm \nu_m^{ssc} < \nu < \nu_{c}^{ssc}}$ & $\frac{4-\epsilon+\beta(8-\epsilon)}{4-\epsilon}$ & $\frac{9-2\epsilon+\beta(\epsilon-2)}{4-\epsilon}$ & 1 & 1.16\% \\
    \hline
    Wind & slow & ${\rm \nu_c^{ssc} < \nu}$ & $\frac{\epsilon-4+\beta(8-\epsilon)}{4-\epsilon}$ & $\frac{6-\epsilon+\beta(\epsilon-2)}{4-\epsilon}$ & 15 & 17.4\% \\
    \hline
    Wind & fast & ${\rm \nu_c^{ssc} < \nu < \nu_{m}^{ssc}}$ & $\frac{\epsilon\beta}{4-\epsilon}$  & $\frac{\epsilon\beta}{4-\epsilon}$  & 2 & 2.33\% \\
    \hline
    Wind & fast & ${\rm \nu_m^{ssc} < \nu }$ & $\frac{\epsilon-4+\beta(8-\epsilon)}{4-\epsilon}$  & $\frac{6-\epsilon+\beta(\epsilon-2)}{4-\epsilon}$  & 15 & 17.4\% \\
    \bottomrule
    \bottomrule
     \end{tabular}
    \label{Table:3}
    \caption{Summary results of CRs without energy injection, showing number and proportion of GRBs satisfying each relation, out of a total of 86 GRBs.}
\end{table}

\begin{figure}[!h]
  \centering
    \subfigure[$\epsilon$= 0.5, slow cooling , ${\rm \nu_m^{ssc} < \nu < \nu_c^{ssc}}$]{\includegraphics[width=0.3\linewidth]{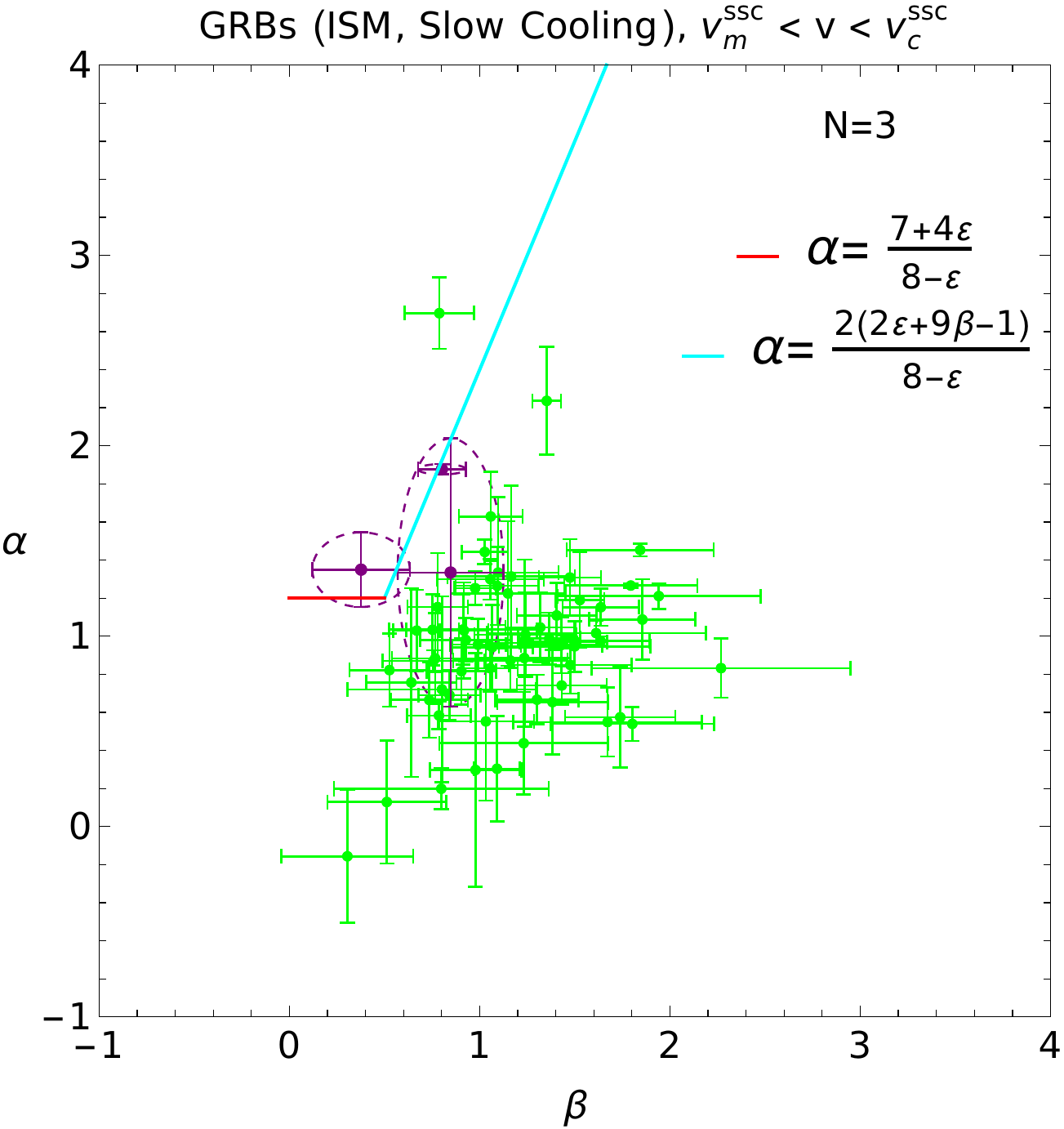}}
    \hfill
    \subfigure[$\epsilon$ = 0.5, fast cooling , ${\rm \nu_c^{ssc} < \nu < \nu_m^{ssc}}$]{\includegraphics[width=0.3\linewidth]{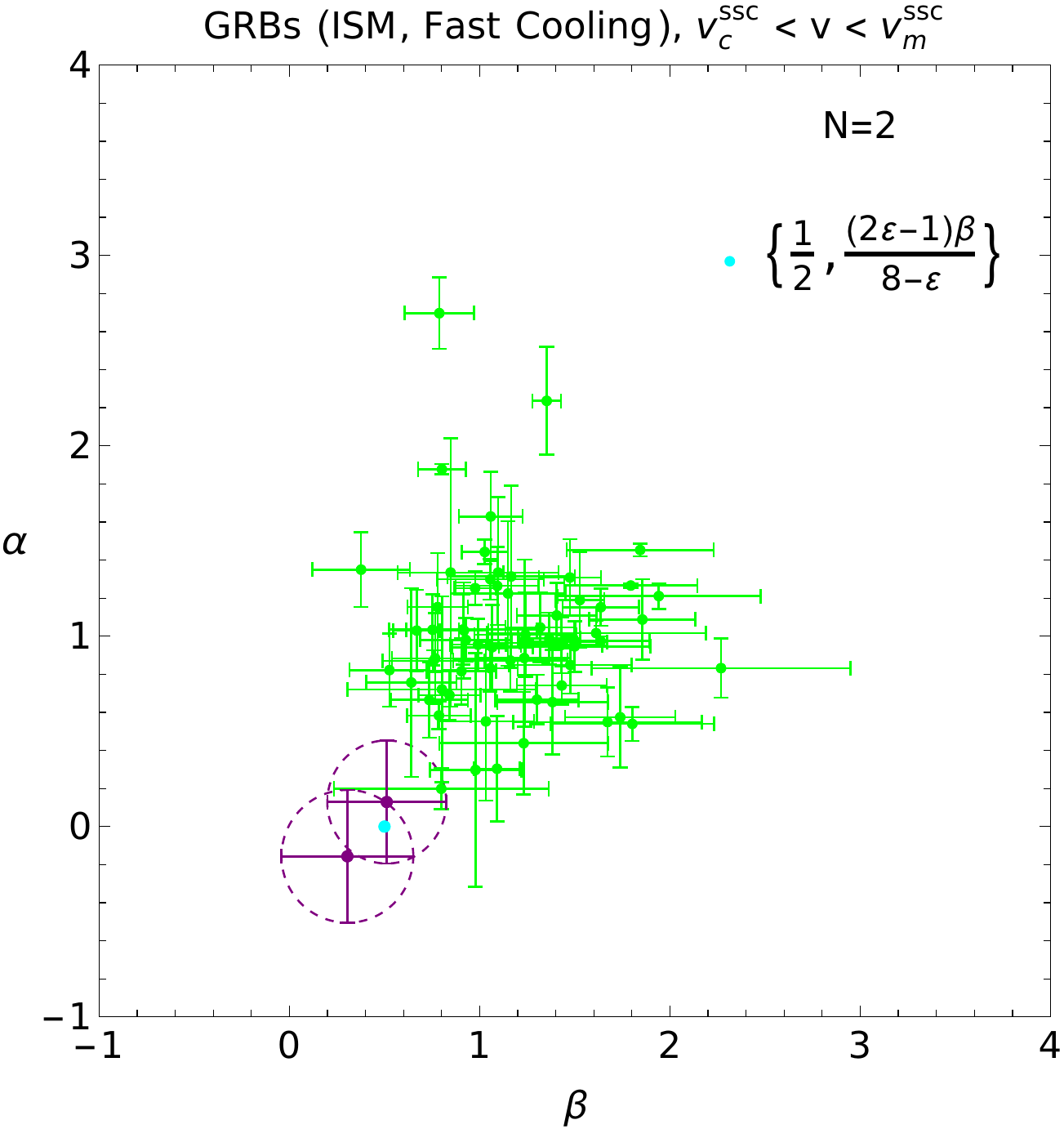}}
    \hfill
    \subfigure[$\epsilon$ = 0.5, ${\rm \nu > max\{\nu_c^{ssc}, \nu_m^{ssc}}\}$]{\includegraphics[width=0.3\linewidth]{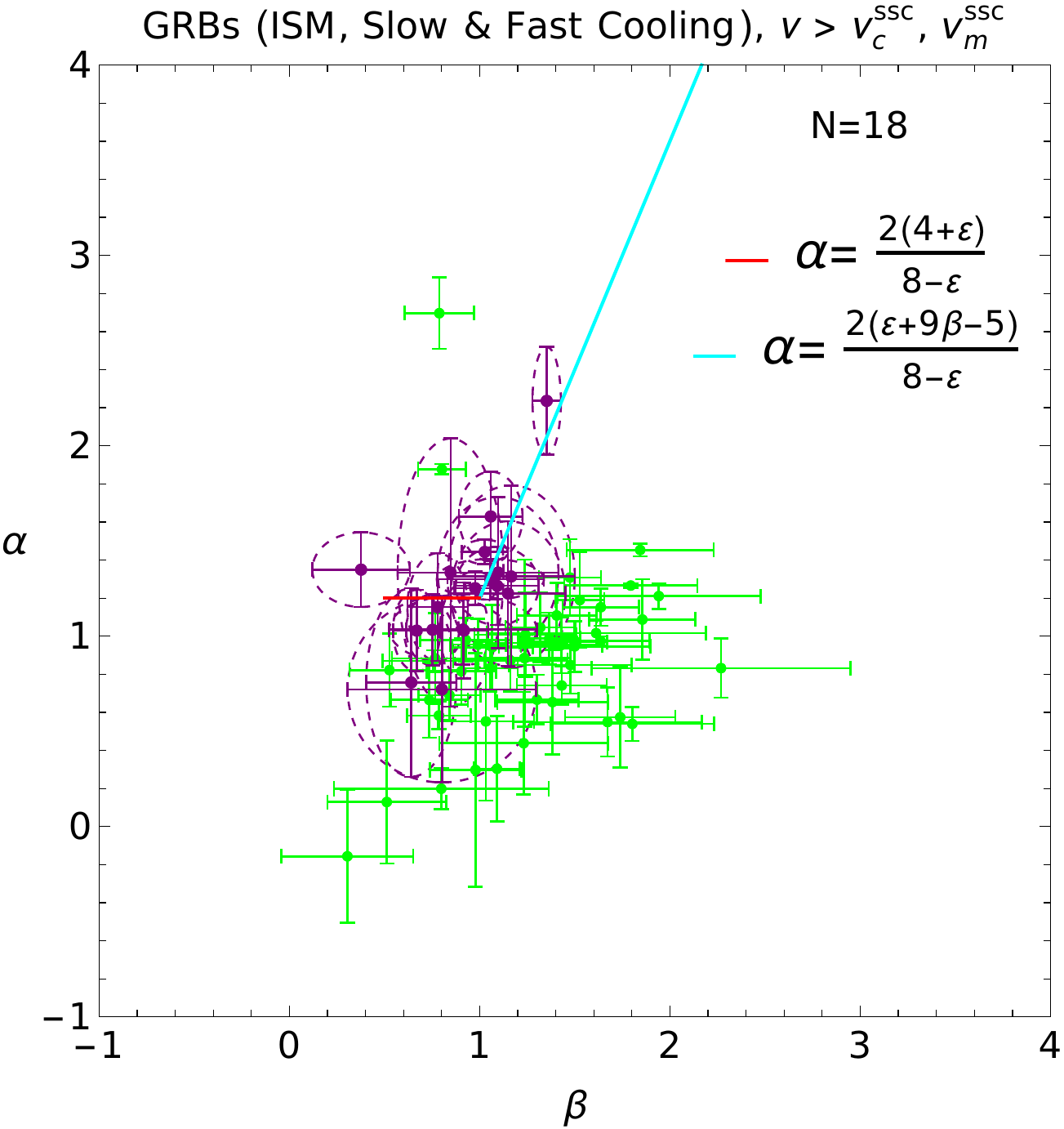}}
    \label{ISM_slow_fast}
    \quad
  \centering
    \subfigure[$\epsilon$ = 0.5, slow cooling , ${\rm \nu_m^{ssc} < \nu < \nu_c^{ssc}}$]{\includegraphics[width=0.3\linewidth]{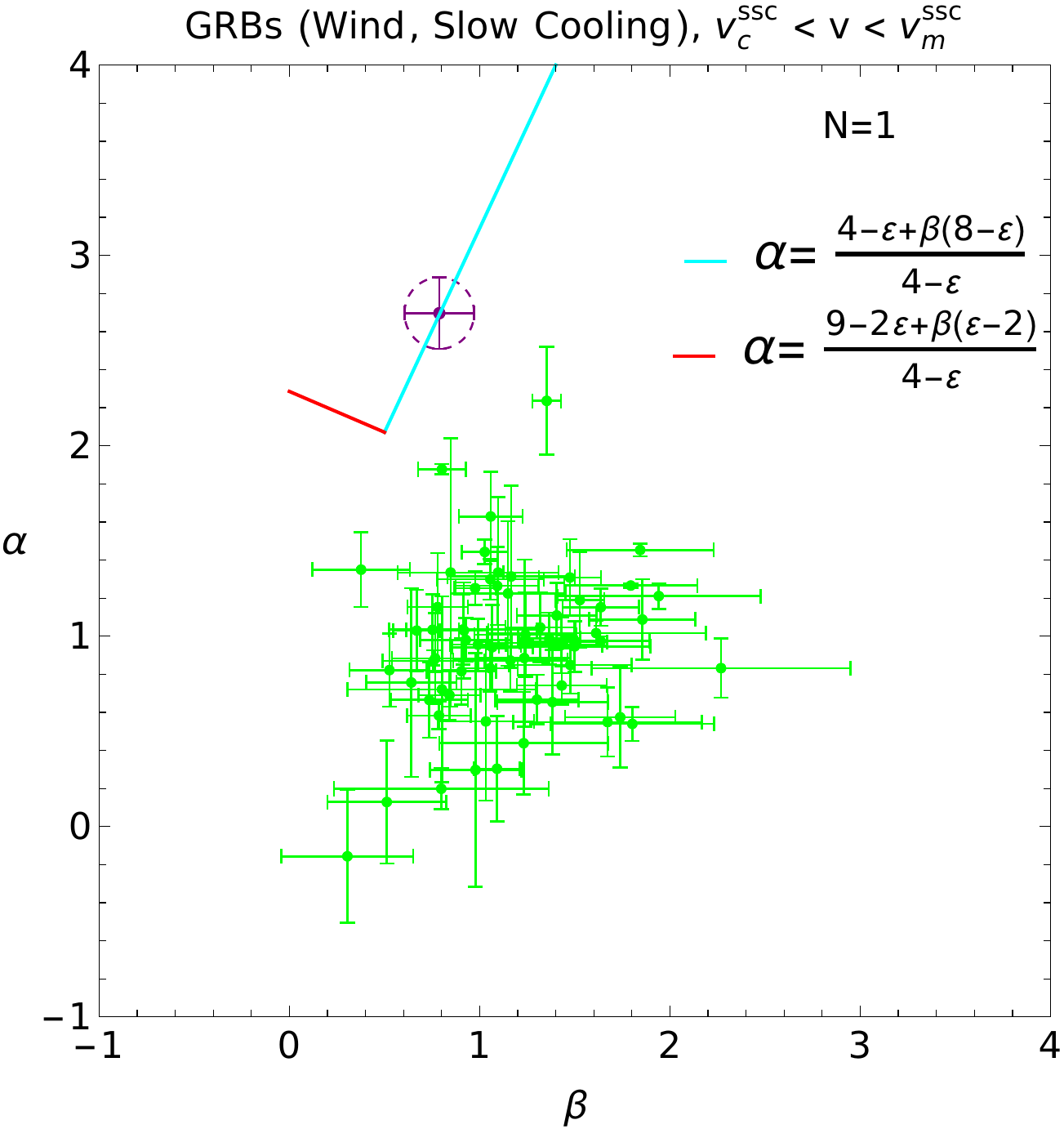}}
    \label{Wind_slow1}
    \hfill
    \subfigure[$\epsilon$= 0.5, fast cooling , ${\rm \nu_c^{ssc} < \nu < \nu_m^{ssc}}$]{\includegraphics[width=0.3\linewidth]{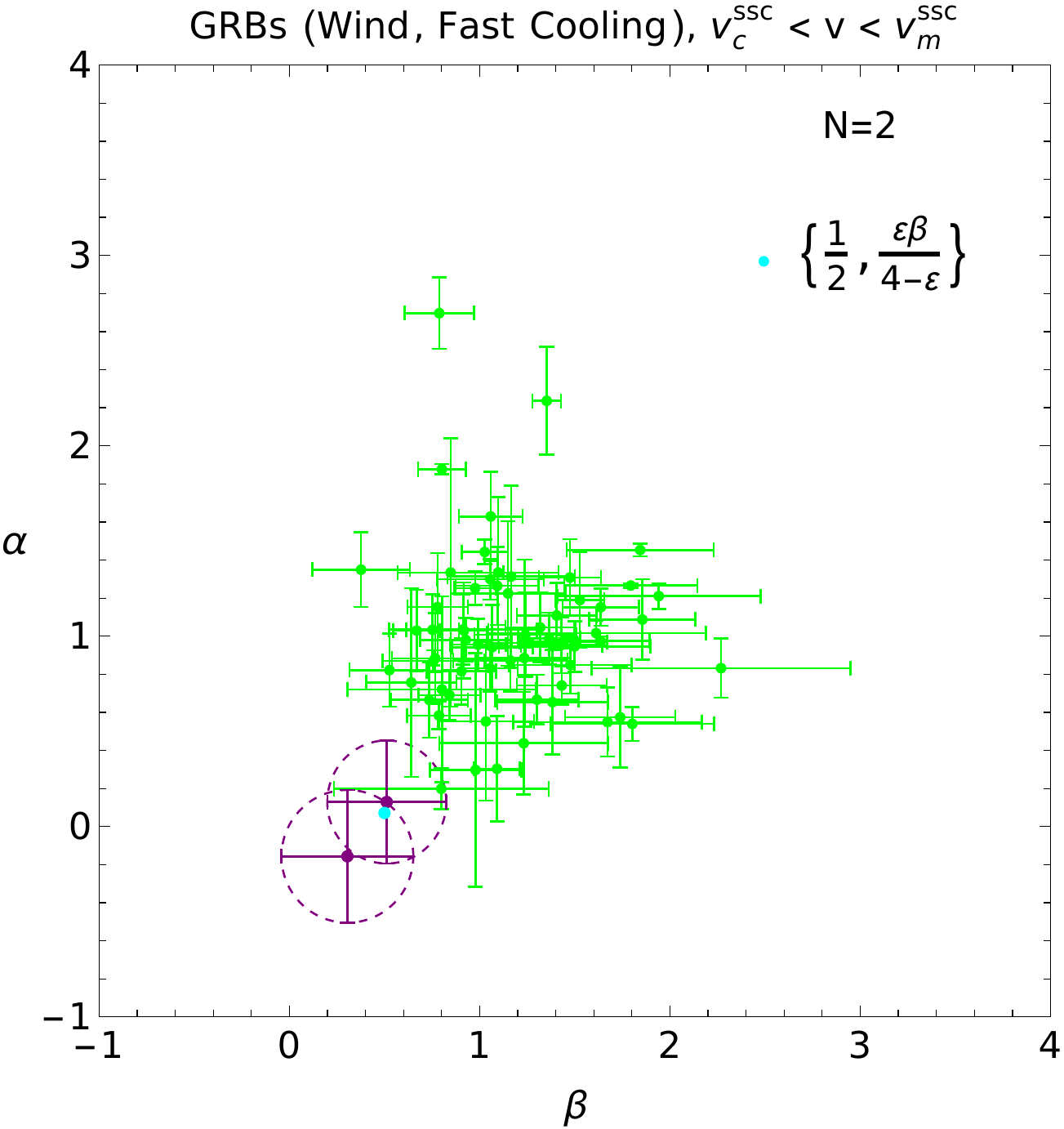}}
    \hfill
    \subfigure[$\epsilon$ = 0.5, ${\rm \nu > max\{\nu_c^{ssc}, \nu_m^{ssc}}\}$]{\includegraphics[width=0.3\linewidth]{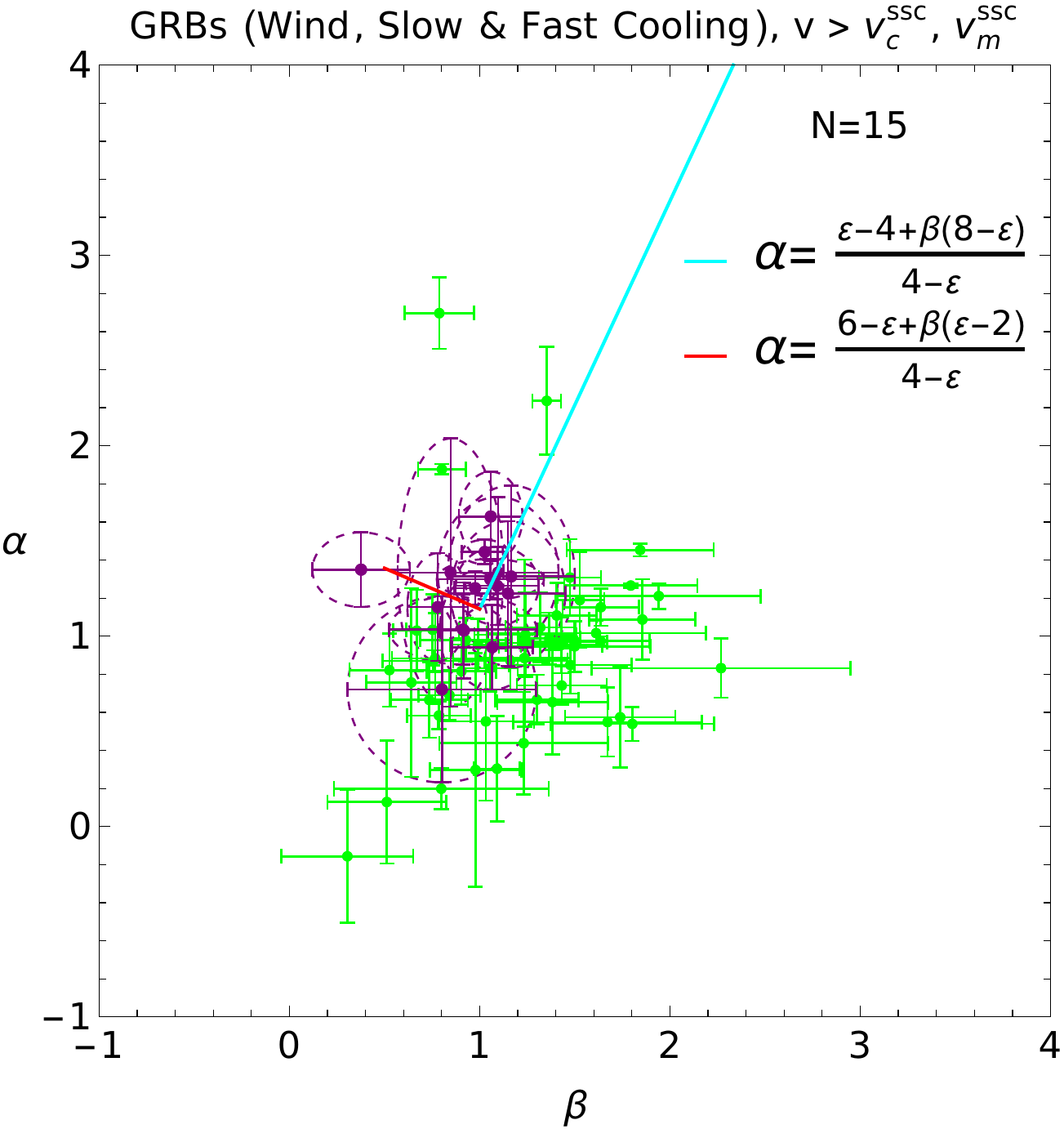}}
    \label{Wind_slow_fast}
    \caption{The panels present the cases of no energy injection for slow and fast cooling regime and in different frequency regimes and with the assumption of $\epsilon$ = 0.5. In each of the subcaption it is detailed the frequency and the regime. The purple ellipses show the cases in which the CR are fulfilled within 1 sigma errorbars. The red, cyan lines and dots correspond to the relationships themselves.}
    \label{no_injection}
\end{figure}

\newpage
%\subsection{Energy injection scenario}

\begin{table}[h]
    \centering\renewcommand{\arraystretch}{1.2}
    \begin{tabular}{c c c c c c c}
    \toprule[1.2pt]
    \toprule[1.2pt]
     Type & Cooling & $\nu$ Range & CR: p $>$ 2  & CR: 1 $<$ p $<$ 2 & GRBs Satisfying Relation & Proportion Satisfying Relation \\
     \midrule
     ISM & slow & ${\rm \nu_m^{ssc} < \nu < \nu_{c}^{ssc}}$ & $\frac{3\beta(q+2)+5q-6}{4}$ & $-$ & 28 & 32.6\% \\
    \hline
    ISM & slow & ${\rm \nu_c^{ssc} < \nu}$ & $\frac{3\beta(q+2)+q-6}{4}$  & $-$ & 29 & 33.7\% \\
    \hline
    ISM & fast & ${\rm \nu_c^{ssc} < \nu < \nu_{m}^{ssc}}$ & $\frac{\beta(5q-6)}{4}$  & $\frac{\beta(5q-6)}{4}$ & 1 & 1.16\% \\
    \hline
    ISM & fast & ${\rm \nu_m^{ssc} < \nu}$ & $\frac{3\beta(q+2)+q-6}{4}$ & $-$ & 29 & 33.7\% \\
    \hline
    Wind & slow & ${\rm \nu_m^{ssc} < \nu < \nu_{c}^{ssc}}$ & $\beta(q+1)+1$ & $\frac{2\beta(q-2)+q+8}{4}$ & 0 & 0\% \\
    \hline
    Wind & slow & ${\rm \nu_c^{ssc} < \nu}$ & $\beta(q+1)-1$ & $\frac{\beta(q-2)+q+2}{2}$  & 30 & 34.9\% \\
    \hline
    Wind & fast & ${\rm \nu_c^{ssc} < \nu < \nu_{m}^{ssc}}$ & $\beta(q-1)$  & $\beta(q-1)$  & 1 & 1.16\% \\
    \hline
    Wind & fast & ${\rm \nu_m^{ssc} < \nu }$ & $\beta(q+1)-1$ & $\frac{\beta(q-2)+q+2}{2}$  & 30 & 34.9\% \\
    \bottomrule
    \bottomrule
    \end{tabular}
    \label{Table:4}
    \caption{Summary results of CRs with energy injection (\textit{q} = 0.5), showing number and proportion of GRBs satisfying each relation, out of a total of 86 GRBs.}
\end{table}

\begin{figure}[!h]
  \centering
    \subfigure[\textit{q} = 0.5, slow cooling , ${\rm \nu_m^{ssc} < \nu < \nu_c^{ssc}}$]{\includegraphics[width=0.3\linewidth]{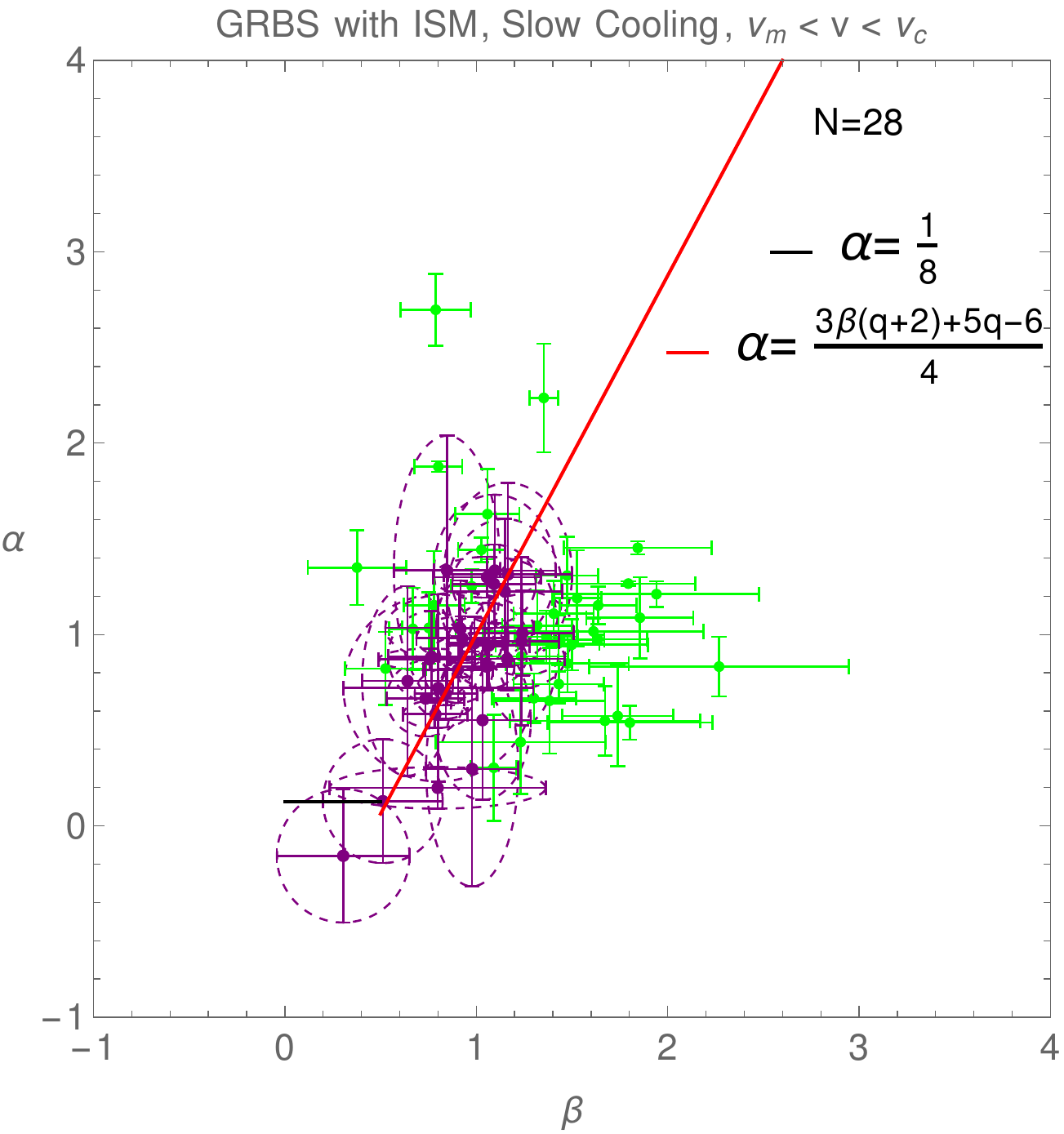}}
    \label{inj_ism_slow}
    \hfill
    \subfigure[\textit{q} = 0.5, fast cooling , ${\rm \nu_c^{ssc} < \nu < \nu_m^{ssc}}$]{\includegraphics[width=0.3\linewidth]{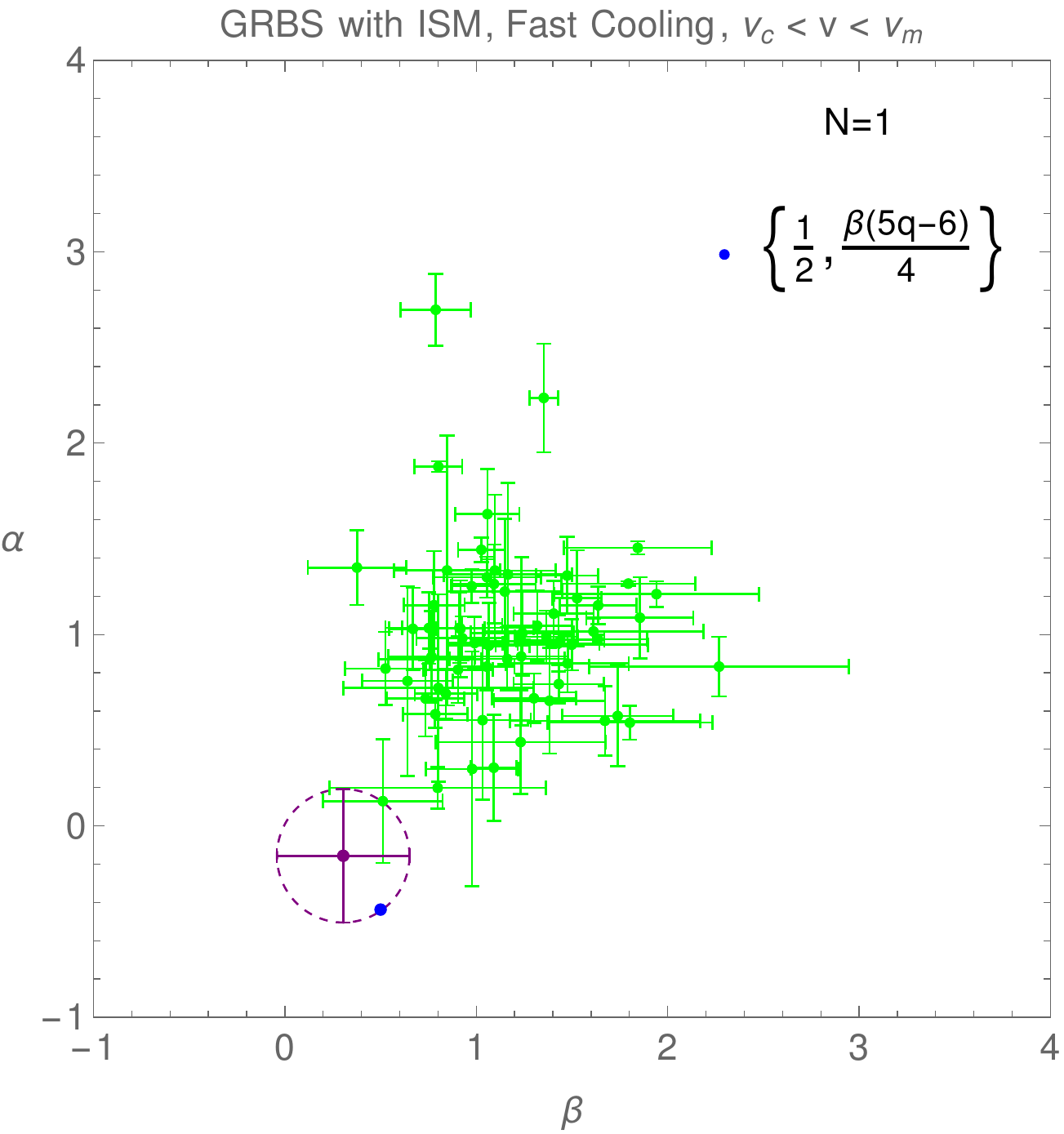}}
    \hfill
    \subfigure[\textit{q}= 0.5, ${\rm \nu > mac\{\nu_c^{ssc}, \nu_m^{ssc}\}}$]{\includegraphics[width=0.3\linewidth]{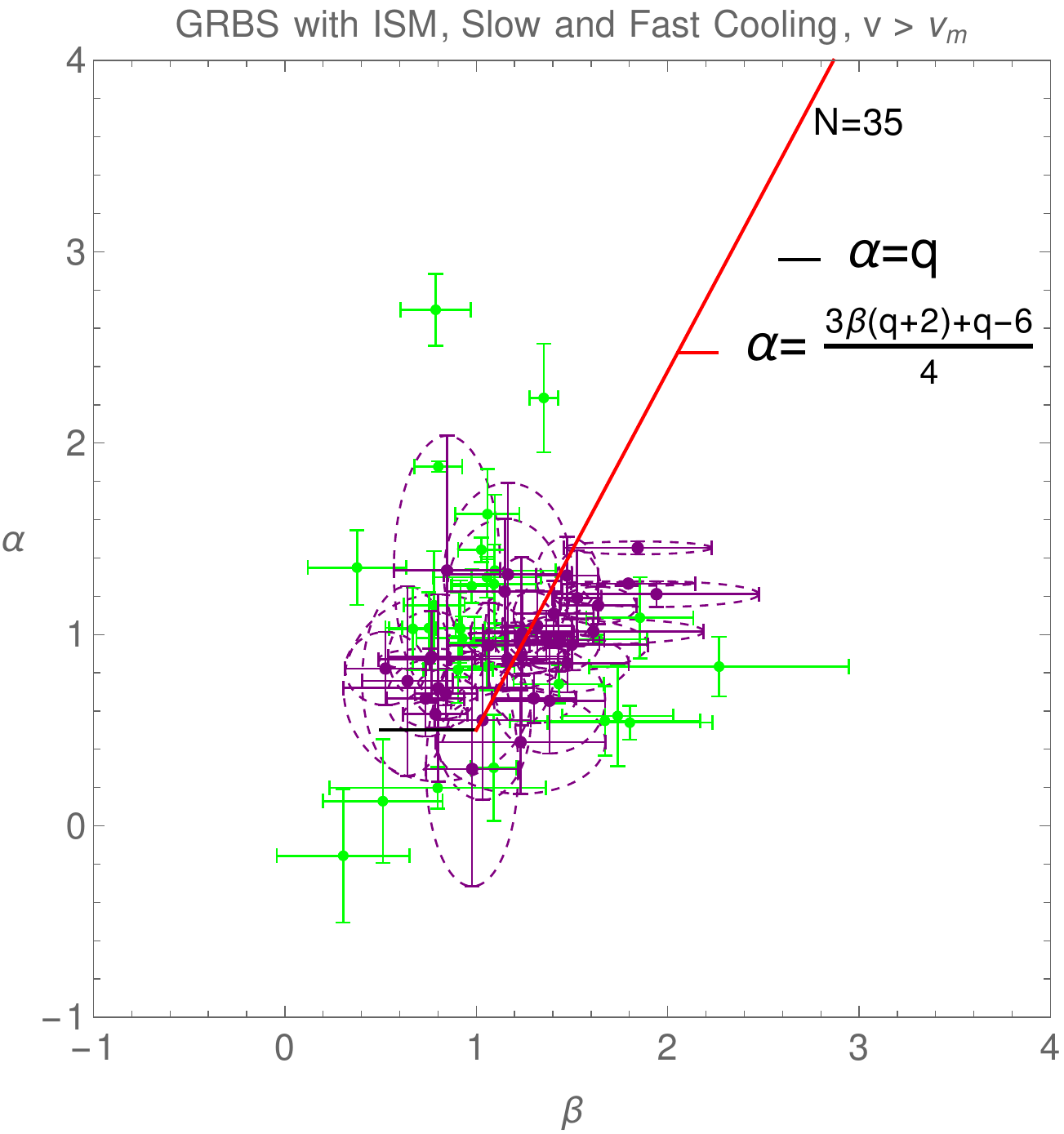}}
    \label{inj_ISM_slow_fast}
    \quad
  \centering
    \subfigure[\textit{q} = 0.5, slow cooling , ${\rm \nu_m^{ssc} < \nu < \nu_c^{ssc}}$]{\includegraphics[width=0.3\linewidth]{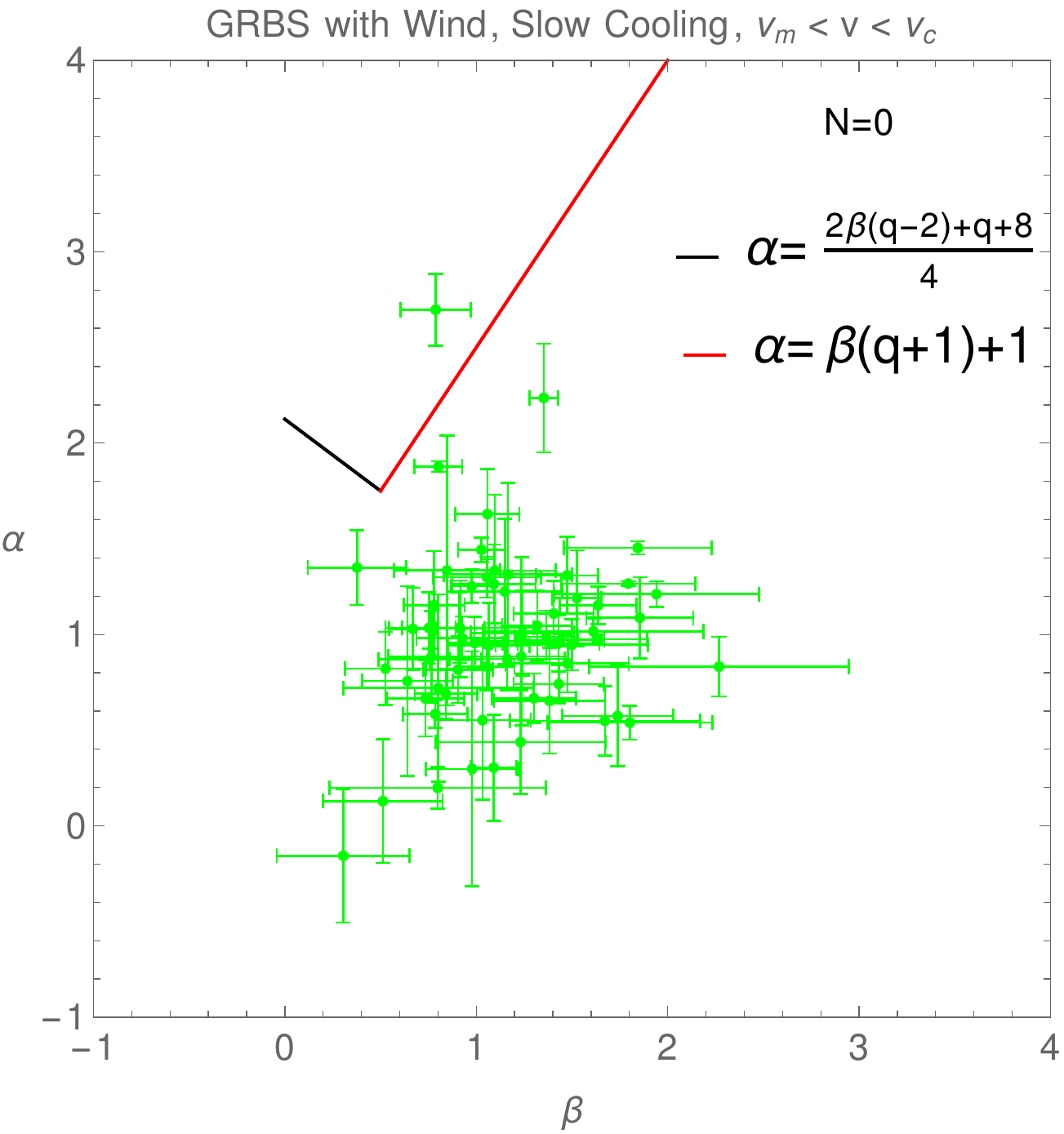}}
    \label{inj_Wind_slow1}
    \hfill
    \subfigure[\textit{q} = 0.5, fast cooling , ${\rm \nu_c^{ssc} < \nu < \nu_m^{ssc}}$]{\includegraphics[width=0.3\linewidth]{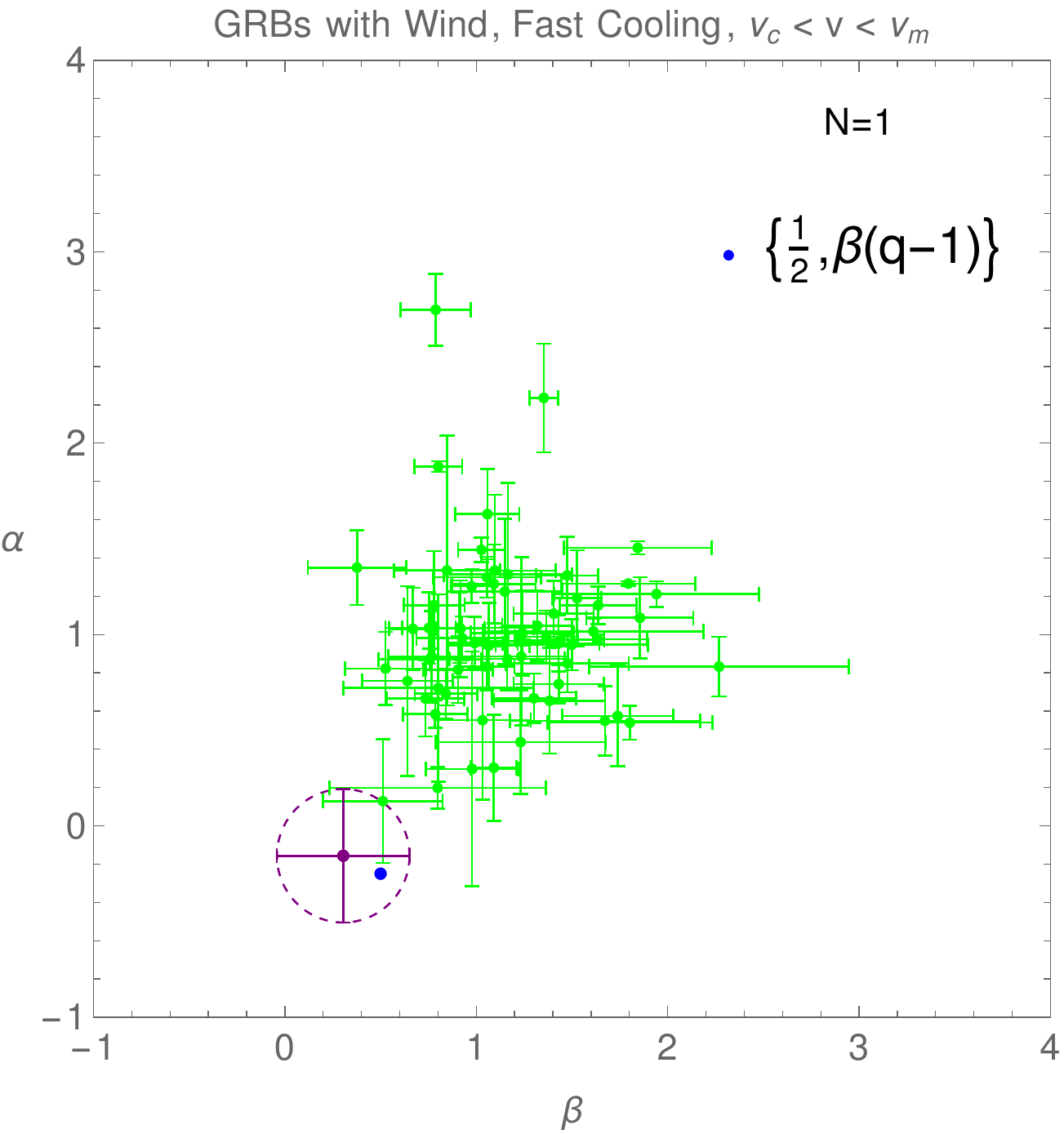}}
    \hfill
    \subfigure[\textit{q} = 0.5, ${\rm \nu > max\{\nu_c^{ssc}, \nu_m^{ssc}\}}$]{\includegraphics[width=0.3\linewidth]{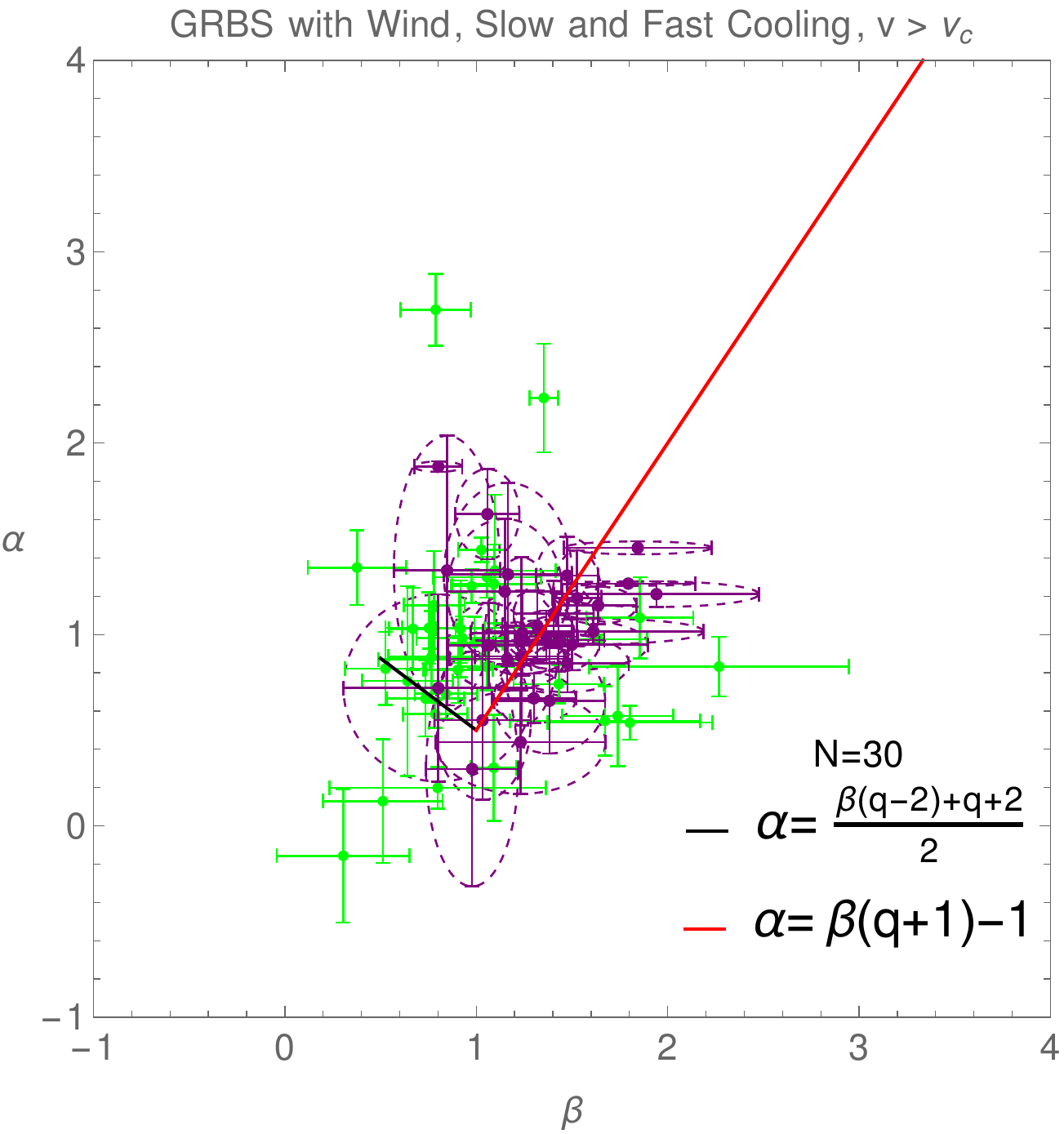}}
    \label{inj_wind_slow_fast}
        \caption{The panels present the cases of energy injection for slow and fast cooling regime and in different frequencies and with the assumption of \textit{q} = 0.5. In each of the subcaption it is detailed the frequency and the regime. Similarly to the previous Fig. the purple ellipses show the cases in which the CR are fulfilled within 1 sigma errorbars. The red, black lines and dots correspond to the relationships themselves.}
    \label{Energy_injection}
\end{figure}

\newpage

\section{Analysis and Discussion}
\label{sec4}

We have derived and listed in Tables \ref{ClosureRelations} and \ref{ClosureRelations2} the CRs of SSC afterglow model in the adiabatic and radiative scenario, and when the central engine continuously injects energy into the blastwave.  We consider the SSC afterglow model evolving in ISM and stellar-wind medium, and the CRs as function of the radiative parameter $\epsilon$, the energy injection index $q$, and the electron spectral index for $1<p<2$ and $ 2\leq p$. The CRs of SSC model in the radiative regime without energy injection and with energy injection evolving in ISM for $1<p<2$ with the cooling conditions of ${\rm \nu^{\rm ssc}_{\rm m}\leq \nu \leq  \nu^{\rm ssc}_{\rm c}}$, and ${\rm \{\nu^{\rm ssc}_{\rm m}, \nu^{\rm ssc}_{\rm c}\}<\nu}$ are not satisfied, and therefore these are not estimated.\\ 

Based on the CRs of SSC afterglow models, we here discuss the results in terms of the bursts reported in 2FLGC.

\subsection{The radiative and the energy injection scenarios}

We begin from the no-energy injection scenario presented in Fig. \ref{no_injection} and \hyperref[Table:3]{Table 3}. Figure \ref{no_injection} shows CRs of SSC model in the radiative regime without energy injection for $\epsilon=0.5$, $1<p<2$ and $p>2$. Panels from (a) to (f) display each cooling condition for slow and fast cooling regime in ISM or stellar-wind medium. \hyperref[Table:3]{Table 3} shows the number and proportion of GRBs satisfying each cooling condition, out of a total of 86 GRBs. The most preferred scenario corresponds to ISM both in the configuration of fast and slow cooling regimes (${\rm max\{\nu_m^{ssc},\nu_c^{ssc} \} < \nu_{\rm LAT}}$, see Fig. \hyperref[ISM_slow_fast]{1(c)})  with a percentage of occurrence of 20.9\% (correspondent to 18 GRBs), followed by the scenario of stellar-wind medium both with slow and fast cooling (${\rm max \{\nu_m^{ssc}, \nu_{c}^{ssc}\} < \nu_{\rm LAT}}$, see Fig. \hyperref[Wind_slow_fast]{1(f)}) with a percentage of occurrence of 17.4\% (correspondent to 15 GRBs). The least fulfilled relation with only one GRB satisfying it is the stellar-wind medium in slow cooling regime (see Fig.\hyperref[Wind_slow1]{1(d)}).  Explicitly, we note that the CRs in the radiative regime could satisfy a fraction of LAT-detected bursts. For instance, the CRs of SSC afterglow model with $\epsilon\approx 1$ and the cooling condition ${\rm max \{\nu^{ssc}_{m}, \nu^{ssc}_{c}\} < \nu_{\rm LAT}}$ favor bursts described with $\beta_{\rm LAT}\sim 1$ and $\alpha_{\rm LAT}\sim 1.5$, and with $0\leq\epsilon\leq 1$ and $\nu^{\rm ssc}_{\rm m}<\nu_{\rm LAT}<\nu^{\rm ssc}_{\rm c}$ favor those burst reported with $\alpha_{\rm LAT}>2$ and $\beta_{\rm LAT}<1$.\\

Figure \ref{Energy_injection} displays CRs of SSC model when the energy is continuously injected into the blastwave for $q=0.5$.   The panels from (a) to (f) present each cooling condition for slow and fast cooling regime in ISM or stellar-wind medium.   \hyperref[Table:4]{Table 4} exhibits the summary results of CRs with energy injection for \textit{q} = 0.5. This table displays the number and proportion of GRBs satisfying each relation, out of a total of 86 GRBs.   The most fulfilled scenario is stellar-wind medium in slow and fast cooling regime (${\rm max\{\nu_m^{\rm ssc}, \nu_{c}^{\rm ssc}\} < \nu_{\rm LAT}}$, see Fig. \hyperref[inj_wind_slow_fast]{2(f)}) with 34.9\% (correspondent to 30 GRBs).  The second most fulfilled scenario is ISM slow and fast cooling regime (${\rm max}\{\nu_m^{\rm ssc}, \nu_{c}^{\rm ssc}\} < \nu_{\rm LAT} $, see Fig. \hyperref[inj_ISM_slow_fast]{2(c)}) with 33.7\% (correspondent to 29 GRBs), followed with a small variation of percentage by the ISM in slow cooling regime ($\nu_m^{\rm ssc} < \nu_{\rm LAT} < \nu_c^{\rm ssc}$, see Fig. \hyperref[inj_ism_slow]{2(a)}) with 32.6\% (correspondent to 28 GRBs).   This shows in this scenario that both ISM and stellar-wind medium are equally possible in the slow cooling regime. Figure \hyperref[inj_Wind_slow1]{2(d)} shows that there is no GRB fulfilled for stellar-wind medium in slow cooling regime. 

%The least fulfilled regime is stellar-wind medium in slow cooling regime (${\rm \nu_m^{ssc} < \nu_{\rm LAT} < \nu_{c}^{ssc}}$) with no GRB fulfilled in Fig. \hyperref[inj_Wind_slow1]{2(d)}.

\subsection{From ISM to stellar-wind environments}

The external environments' density profiles cover short and long GRB progenitors are used.   ISM (${\rm k}=0$) is associated with the merger of binary compact objects; a black hole (BH) - a neutron star (NS) or NS-NS \citep{1992ApJ...392L...9D, 1992Natur.357..472U, 1994MNRAS.270..480T, 2011MNRAS.413.2031M}, which is the progenitor of short GRBs, and the stellar-wind medium ($k=2$) is related with the death of massive stars with different mass-loss evolution, which is the progenitor of long GRBs. It is worth noting that the core-collapse (CC) of dying massive stars could be associated with ISM \citep{1993ApJ...405..273W,1998ApJ...494L..45P} when the wind ejected by its progenitor does not extend beyond the deceleration ratio and a wind-to-ISM afterglow transition could be expected in the scenario of CC of massive stars \citep[e.g., see][]{2017ApJ...848...15F,2019ApJ...879L..26F}. We find that 23 GRBs within ISM and 18 with stellar-wind medium satisfy at least one CR in the scenario without injection. 45 GRBs do not follow any of the CR. For the injection scenario, 28 GRBs do not satisfy any CRs and 58 satisfy at least one CR.

The bright and hard GRB 090510 is the only short burst in our subset with energies above 10 GeV. A photon energy of $30\,{\rm GeV}$ which cannot be described with synchrotron afterglow model was detected at 0.8 s after the trigger time from GRB 090510 \citep{2010ApJ...716.1178A}. This burst associated with a merger of two NSs as progenitor is naturally interpreted as a superposition of synchrotron and SSC afterglow model involving in ISM \citep[without energy injection][]{2011ApJ...733...22H, 2016ApJ...831...22F}, which is consistent with our results.   

The long GRB 130427A is the more energetic burst with a redshift $z<1$ in our sample with clear evidence of the SSC scenario. From this bright burst was detected 14 photons with energies above 10 GeV, including a 95-GeV and a 32-GeV photon observed a few minutes and after more than 9 hours after the trigger time, respectively. The highest energy photons cannot be interpreted in the synchrotron afterglow model \citep[e.g., see][]{2016ApJ...818..190F}.

\cite{2013ApJ...779L...1K} analyzed the multi-wavelength observations of GRB 130427A and modeled the observations below GeV energy range with a synchrotron afterglow model in the adiabatic regime without energy injection for $p>2$. The authors found that an intermediate value of density profile between ${\rm k}=0$ and ${\rm k}=2$ was consistent with these observations, suggesting that the massive progenitor could have produced an eruption before the CC event, altering the external density profile. To analyze GRB 130427A, we derive the SSC afterglow model for ${\rm k}$ in general in the adiabatic regime ($\epsilon=0$) and with energy injection ($q=1$).   In this case,  the spectral breaks and the maximum flux of SSC scenario evolve as $\nu^{\rm ssc}_{\rm m}\propto t^{-\frac{18-5k}{2(4-k)}}$, $\nu^{\rm ssc}_{\rm c}\propto t^{-\frac{2-5k}{2(4-k)}}$ and $F^{\rm ssc}_{\rm max}\propto t^{\frac{2-3k}{2(4-k)}}$, respectively. Therefore, following the procedure shown in section \ref{sec2}, we can write the CRs in the fast and slow cooling regime for $p>2$ as

{\small
\begin{eqnarray}
\label{ssc_ism}
F^{\rm ssc}_{\nu}\propto  \cases{ 
t^{\frac{2-k}{4(4-k)}} \nu^{-\frac12}, \hspace{2.6cm}  \nu^{\rm ssc}_{\rm c}<\nu <\nu^{\rm ssc}_{\rm m},\hspace{.25cm}\cr
t^{\frac{2(10-3k)+p(5k-18)}{4(4-k)}}\,\nu^{-\frac{p}{2}},\,\,\,\, \hspace{0.9cm}  \nu^{\rm ssc}_{\rm m} <\nu\,, \cr
}
\end{eqnarray}
}
and
{\small
\begin{eqnarray}
\label{ssc_ism}
F^{\rm ssc}_{\nu}\propto  \cases{ 
t^{\frac{11(2-k)+p(5k-18)}{4(4-k)}} \nu^{-\frac{p-1}{2}}, \hspace{1.3cm} \nu^{\rm ssc}_{\rm m}<\nu <\nu^{\rm ssc}_{\rm c},\hspace{.25cm}\cr
t^{\frac{2(10-3k)+p(5k-18)}{4(4-k)}}\,\nu^{-\frac{p}{2}},\,\,\,\, \hspace{1.2cm}  \nu^{\rm ssc}_{\rm c} <\nu\,, \cr
}
\end{eqnarray}
}
respectively. Considering the spectral and temporal indexes reported in 2FLGC ($\beta=1.12\pm0.6$ and $\alpha=1.24\pm0.6$), we can notice that with $p=2.2$, the CRs are satisfied in the cooling condition ${\rm max\{ \nu^{\rm ssc}_{\rm m}, \nu^{\rm ssc}_{\rm c} \} <\nu}$ for any value of ${\rm k}$.

\subsection{Synchrotron forward-shock limit}

The maximum photon energy radiated by the synchrotron process in ISM and stellar-wind medium without and with energy injection can be written as

\begin{eqnarray}
\label{ene_max_eps}
h\nu^{\rm syn}_{\rm max} \approx \cases{ 
1.0\, {\rm GeV}\left(\frac{1+z}{2}\right)^{\frac{\epsilon-5}{8-\epsilon}}\,  n^{-\frac{1}{8-\epsilon}}E_{53}^{\frac{1}{8-\epsilon}}\Gamma_{0,2.7}^{-\frac{\epsilon}{8-\epsilon}}t_{2}^{-\frac{3}{8-\epsilon}} \hspace{1.3cm} {\rm for} \hspace{0.2cm} {\rm k=0 }\cr
0.5\, {\rm GeV}\left(\frac{1+z}{2}\right)^{\frac{\epsilon-3}{4-\epsilon}}\,    A_{\rm W,-1}^{-\frac{1}{4-\epsilon}}E_{53}^{\frac{1}{4-\epsilon}}\Gamma_{0,2.7}^{-\frac{\epsilon}{4-\epsilon}}t_{2}^{-\frac{1}{4-\epsilon}} \hspace{1.3cm} {\rm for} \hspace{0.2cm} {\rm k=2}\,,
}
\end{eqnarray}

and 

\begin{eqnarray}
\label{ene_max_q}
h\nu^{\rm syn}_{\rm max} \approx \cases{ 
1.1\, {\rm GeV}\left(\frac{1+z}{2}\right)^{-\frac{5}{8}}\,    n^{-\frac{1}{8}}E_{53}^{\frac{1}{8}}\,t_{2}^{-\frac{2+q}{8}} \hspace{1.8cm} {\rm for} \hspace{0.2cm} {\rm k=0 }\cr
0.7\, {\rm GeV}\left(\frac{1+z}{2}\right)^{-\frac{3}{4}}\,    A_{\rm W,-1}^{-\frac{1}{4}}E_{53}^{\frac{1}{4}}\,t_{2}^{-\frac{q}{4}} \hspace{1.7cm} {\rm for} \hspace{0.2cm} {\rm k=2}\,,
}
\end{eqnarray}

respectively.  The synchrotron scenario cannot explain photons with energies above the synchrotron limit, although the temporal and spectral indexes of GRBs detected by Fermi-LAT can be described in this scenario. Therefore, the synchrotron scenario is not the appropriate model, and a new mechanism such as SSC would be the most indicated.   The photo-hadronic interactions could emit energetic photons beyond the synchrotron limit \citep[e.g., see][]{1997PhRvL..78.2292W}, however the non coincidences of neutrinos with GRBs reported by the IceCube collaboration constrain dramatically the amount of hadrons involved in these interactions \citep{2012Natur.484..351A, 2016ApJ...824..115A, 2015ApJ...805L...5A}. Therefore, the  LAT  photons below the synchrotron limit might be explained in the synchrotron scenario, and beyond this limit, SSC would explain the LAT photons. For instance,  Table 7 in 2FLGC exhibits 29 Fermi-LAT bursts with photon energies above $> 10\,{\rm GeV}$, which would have to satisfy the CRs of SSC afterglow scenarios, at least in a short time interval. It is worth noting that the maximum energy of the SSC process  is much larger than the LAT energy range ($h\nu^{\rm ssc}_{\rm max}=\gamma^2_{\rm max} h\nu^{\rm syn}_{\rm max}\gg 300\,{\rm GeV}$).

%slow cooling regime
\subsection{The cooling conditions and microphysical parameter $\epsilon_B$}

\cite{2019ApJ...883..134T} considered temporal and spectral indexes and did a systematic analysis of the closure relations in a sample of 59 selected LAT-detected bursts. On the one hand, they found that although in most cases the standard synchrotron emission describes the spectral and temporal indexes, there is still a considerable fraction of bursts that can hardly be described with this model. On the other hand,  they found that several GRBs satisfy the closure relations of the slow-cooling regime ($\nu^{\rm syn}_{\rm m}<\nu_{\rm LAT}<\nu^{\rm syn}_{\rm c})$, but when the magnetic microphysical parameter has an atypical small value of $\epsilon_B<10^{-7}$.  It is worth noting that several afterglow modelling studies have revealed values in the range of $3.5\times10^{-5}\leq\epsilon_B\leq0.33$ \citep[e.g., see][]{2014ApJ...785...29S}, and not as small as those reported by \cite{2019ApJ...883..134T}.\\ 

Here, we constrain the value of $\epsilon_B$ that allows the Fermi-LAT observations to lie in the cooling condition $\nu^{\rm ssc}_{\rm m}<\nu_{\rm LAT}<\nu^{\rm ssc}_{\rm c}$ or $\nu^{\rm ssc}_{\rm m}<\nu^{\rm ssc}_{\rm c}<\nu_{\rm LAT}$ for typical values of GRB afterglows. Considering cooling spectral break $\nu^{\rm ssc}_{\rm c}=\nu_{\rm LAT}$ in Eqs. \ref{ssc_br_hom_v1}, \ref{ssc_br_win_eps}, \ref{ssc_br_hom_q} and \ref{ssc_br_win}, the microphysical parameter in the scenario without and with energy injection for ISM and stellar-wind medium is

{\small
\begin{eqnarray}
\label{eps_B1}
\epsilon_B \lesssim \cases{ 
2.5\times10^{-4}\left(\frac{1+z}{2}\right)^{-\frac{6(2+\epsilon)}{7(8-\epsilon)}}\, (1+Y(\gamma_c))^{-\frac87}  n_{-1}^{\frac{7\epsilon-36}{7(8-\epsilon)}}E_{53}^{-\frac{20}{7(8-\epsilon)}}\Gamma_{0,2.7}^{\frac{20\epsilon}{7(8-\epsilon)}}t_{3}^{\frac{4(2\epsilon-1)}{7(8-\epsilon)}} \left(\frac{h\nu^{\rm ssc}_c}{100\,{\rm MeV}}\right)^{-\frac27} \hspace{1.3cm} {\rm for} \hspace{0.2cm} {\rm k=0 }\cr
1.7\times10^{-5}\left(\frac{1+z}{2}\right)^{\frac{8(\epsilon-3)}{7(4-\epsilon)}}\, (1+Y(\gamma_c))^{-\frac87}  A_{W,-1}^{\frac{7\epsilon-36}{7(4-\epsilon)}}E_{53}^{\frac{8}{7(4-\epsilon)}}\Gamma_{0,2.7}^{-\frac{8\epsilon}{7(4-\epsilon)}}t_{3}^{\frac{2(8-3\epsilon)}{7(4-\epsilon)}} \left(\frac{h\nu^{\rm ssc}_c}{100\,{\rm MeV}}\right)^{-\frac27} \hspace{1.3cm} {\rm for} \hspace{0.2cm} {\rm k=2 }
}
\end{eqnarray}
}

and

{\small
\begin{eqnarray}
\label{eps_B2}
\epsilon_B \lesssim  \cases{ 
2.2\times10^{-4}\left(\frac{1+z}{2}\right)^{-\frac{3}{14}}\, (1+Y(\gamma_c))^{-\frac87}  n_{-1}^{-\frac{9}{14}}E_{53}^{-\frac{5}{14}}t_{3}^{\frac{5q-6}{14}} \left(\frac{h\nu^{\rm ssc}_c}{100\,{\rm MeV}}\right)^{-\frac27} \hspace{1.3cm} {\rm for} \hspace{0.2cm} {\rm k=0 }\cr
6.3\times10^{-5}\left(\frac{1+z}{2}\right)^{-\frac{6}{7}}\, (1+Y(\gamma_c))^{-\frac87}  A_{W,-1}^{-\frac{9}{7}}E_{53}^{\frac{2}{7}}t_{3}^{\frac{2(3-q)}{7}} \left(\frac{h\nu^{\rm ssc}_c}{100\,{\rm MeV}}\right)^{-\frac27} \hspace{1.3cm} {\rm for} \hspace{0.2cm} {\rm k=2 }\,,
}
\end{eqnarray}
}
respectively. When the Klein - Nishina (KN) effects are not negligible, the value of the Compton parameter is not constant and then defined as

{\small
\begin{eqnarray}
\label{Yth_kn}
Y(\gamma_c)[Y(\gamma_c)+1] = \frac{\epsilon_{e}}{\epsilon_{B}}  \left(\frac{\gamma_{\rm c}}{\gamma_{\rm m}}\right)^{2-p}\,\cases{ 
\left( \frac{\nu^{\rm syn}_{\rm m}}{\nu^{\rm syn}_{\rm c}} \right)^{-\frac{p-3}{2}}\,\left( \frac{\nu^{\rm syn}_{\rm KN}(\gamma_{\rm c})}{\nu^{\rm syn}_{\rm m}}\right)^\frac43 \, \hspace{1cm} {\rm for} \hspace{0.2cm} {\nu^{\rm syn}_{\rm KN}(\gamma_{\rm c}) <\nu^{\rm syn}_{\rm m} }\cr
\left(\frac{\nu^{\rm syn}_{\rm KN}(\gamma_{\rm c})}{\nu^{\rm syn}_c}\right)^{-\frac{p-3}{2}} \, \hspace{2.2cm} {\rm for} \hspace{0.2cm} { \nu^{\rm syn}_{\rm m} < \nu^{\rm syn}_{\rm KN}(\gamma_{\rm c}) < \nu^{\rm syn}_{\rm c}}\cr
1\, \hspace{4.25cm} {\rm for} \hspace{0.2cm} {\nu^{\rm syn}_{\rm c} < \nu^{\rm syn}_{\rm KN}(\gamma_{\rm c})}\,.
%&& \hspace{0.1cm}\,     \hspace{0.8cm} \cr
%\hspace{4.05cm} {\rm for} \hspace{0.2cm} {2 \leq p }\,,
}
\end{eqnarray}
}

where $h\nu^{\rm syn}_{\rm KN} (\gamma_c) \simeq\frac{2\Gamma}{(1+z)}\,\frac{m_e c^2}{\gamma_c}$ for $\nu^{\rm syn}_{\rm m} < \nu^{\rm syn}_{\rm c}$ \citep[see][]{2009ApJ...703..675N, 2010ApJ...712.1232W}.   To describe the LAT observations above 100 MeV, it is needed to define the Lorentz factor of those electrons ($\gamma_{*}$) that might emit high-energy photons via synchrotron process, and a new spectral break $h\nu^{\rm syn}_{\rm KN}(\gamma_*)$ would have to be included and the Compton parameter $Y(\gamma_{*})$ re-computed.  For instance, the new Compton parameter becomes {\small $Y(\gamma_*)=Y(\gamma_c) \left(\frac{\nu_{*}}{\nu_c}\right)^{\frac{p-3}{4}}   \left(\frac{\nu^{\rm syn}_{\rm KN}(\gamma_{\rm c})}{\nu^{\rm syn}_c}\right)^{-\frac{p-3}{2}}$} for $ \nu^{\rm syn}_{\rm m} <  h\nu^{\rm syn}_{\rm KN}(\gamma_*)=100\,{\rm MeV} < \nu^{\rm syn}_{\rm c} < \nu^{\rm syn}_{\rm KN}(\gamma_{\rm c})$  \citep[for details, see][]{2010ApJ...712.1232W}.\\

We can notice that for $Y(\gamma_c)\ll 1$, Eqs. \ref{eps_B1} and \ref{eps_B2} are not altered, and  for $Y(\gamma_c)\gg 1$, $\epsilon_B$ increases. For instance,  for  $Y(\gamma_c)\gg 1$ and with the spectral breaks in the hierachy $ \nu^{\rm syn}_{\rm m} <  h\nu^{\rm syn}_{\rm KN}(\gamma_*) < \nu^{\rm syn}_{\rm c} < \nu^{\rm syn}_{\rm KN}(\gamma_{\rm c})$, the new Compton parameter becomes $Y(\gamma_*)\simeq 1$ for typical values of GRB afterglow \citep[see ][]{2010ApJ...712.1232W}. In all cases, $\epsilon_B$ lies in the range of values reported in \cite{2014ApJ...785...29S} when SSC afterglow model is considered.

\section{conclusion}
\label{sec5}

%%%%%%%%%%%%%%%%
We have derived the CRs of the SSC afterglow model in the adiabatic and radiative scenario, and when the central engine continuously injects energy into the blastwave to study the evolution of the spectral and temporal indexes of bursts reported in 2FLGC, which comprises a subset of 29 bursts with photon energies above a few GeV. We consider the SSC afterglow model evolving in ISM (${\rm k=0}$) and stellar-wind medium (${\rm k=2}$) and the CRs as function of the radiative parameter $\epsilon$, the energy injection index $q$, and the electron spectral index for $1<p<2$ and $ 2\leq p$. The CRs of SSC model in the radiative regime (without energy injection) and with energy injection evolving in ISM for $1<p<2$ with the cooling conditions of $\nu^{\rm ssc}_{\rm m}\leq \nu \leq  \nu^{\rm ssc}_{\rm c}$, and $\{\nu^{\rm ssc}_{\rm m}, \nu^{\rm ssc}_{\rm c}\}<\nu$ are not satisfied, and therefore these are not estimated. Based on the description of the multi-wavelength observations of GRB 130427A with a density profile index between ISM and stellar-wind medium, we have derived the CRs of SSC  adiabatic model without energy injection  for ${\rm k}$ in general.  
%%%%%%%%%%%%%%%%%

After analyzing the bursts reported in 2FLGC, we have shown that: i) in the radiative model without energy injection, the most preferred scenario corresponds to ISM in fast and slow cooling regimes (${\rm max}\{\nu_m^{\rm ssc}, \nu_{c}^{\rm ssc}\} < \nu_{\rm LAT} $) with a percentage of occurrence of 20.9\% (correspondent to 18 GRBs), followed by the scenario of stellar-wind medium both in slow and fast cooling (${\rm max}\{\nu_m^{\rm ssc}, \nu_{c}^{\rm ssc}\} < \nu_{\rm LAT} $) with a percentage of occurrence of 17.4\% (correspondent to 15 GRBs). The least fulfilled relation with only one GRB satisfying it is the stellar-wind medium in a slow cooling regime.   ii) in the SSC model with energy injection, the most fulfilled scenario is stellar-wind environment in slow and fast cooling regime ($\nu_{\rm LAT} > {\rm max}\{\nu_{c}^{\rm ssc}, \nu_m^{\rm ssc}\}$) with 34.9\% (correspondent to 30 GRBs).  The second most fulfilled scenario is the ISM in a slow and fast cooling regime (${\rm max}\{\nu_m^{\rm ssc}, \nu_{c}^{\rm ssc}\} < \nu_{\rm LAT}$) with 33.7\% (correspondent to 29 GRBs), followed with a slight variation of percentage by ISM in slow cooling regime ($\nu_m^{\rm ssc} < \nu_{\rm LAT} < \nu_c^{\rm ssc}$) with 32.6\% (correspondent to 28 GRBs). It shows that both ISM and wind are equally possible in the slow cooling regime. The least fulfilled regime is the stellar wind medium in slow cooling regime ($\nu_m^{ssc} < \nu_{\rm LAT} < \nu_{c}^{ssc}$) with no GRB fulfilled.\\

\cite{2010MNRAS.403..926G} analyzed the high-energy emission of a sample of GRBs detected by the Fermi-LAT instrument. The authors reported that the LAT emission evolved as $\sim t^{-\frac{10}{7}}$ instead of the standard synchrotron model ($\sim t^{-1}$), suggesting that the synchrotron afterglow model lay in the fully radiative regime.  Therefore, as in this work, it is natural to assume that SSC emission also lies in the radiative regime. Explicitly, we note that the CRs in the radiative regime could also satisfy a fraction of bursts reported in 2FLGC.  For instance, the CRs with $\epsilon\approx 1$ and the cooling condition ${\rm max \{\nu^{ssc}_{m}, \nu^{ssc}_{c}\} < \nu_{\rm LAT}}$ favor bursts described with $\beta_{\rm LAT}\sim 1$ and $\alpha_{\rm LAT}\sim 1.5$, and the CRs with $0\leq\epsilon\leq 1$ and $\nu^{\rm ssc}_{\rm m}<\nu_{\rm LAT}<\nu^{\rm ssc}_{\rm c}$ favor those burst reported with $\alpha_{\rm LAT}>2$ and $\beta_{\rm LAT}<1$.\\

The standard synchrotron forward-shock model successfully explains the GRB afterglow observations and the evolution of spectral and temporal indexes below the synchrotron limit. However, a sample of 29 bursts in 2FLGC has been detected with photon energies above $> 10 GeV$, which can hardly be described in the synchrotron scenario. Based on the maximum photon energy emitted by synchrotron radiation during the deceleration phase being around $\approx 1 \, {\rm GeV}$, we claim that this sample of bursts cannot be adequately interpreted in terms of the synchrotron model, but SSC afterglow model. We discard the photo-hadronic interactions because of the non-coincidences of neutrinos with GRBs reported by the IceCube collaboration \citep{2012Natur.484..351A, 2016ApJ...824..115A, 2015ApJ...805L...5A}. Consequently, although the CRs of the synchrotron standard model could satisfy this sample of 29 bursts,  the synchrotron model is not the appropriate one, and the SSC process could be the most suitable.\\

\cite{2019ApJ...883..134T} considered temporal and spectral indexes and systematically analyzed the closure relations in a sample of 59 selected LAT-detected bursts. On the one hand, they found that although the standard synchrotron emission describes the spectral and temporal indexes in most cases, there is still a considerable fraction of bursts that can hardly be described with this model. On the other hand,  they found that several GRBs satisfy the closure relations of the slow-cooling regime ($\nu^{\rm syn}_{\rm m}<\nu_{\rm LAT}<\nu^{\rm syn}_{\rm c})$, but when the magnetic microphysical parameter has an atypical small value of $\epsilon_B<10^{-7}$.  In this work,  we have shown that a typical value of $\epsilon_B$ lying in the range of $3.5\times10^{-5}\leq\epsilon_B\leq0.33$ \citep[e.g., see][]{2014ApJ...785...29S} is obtained when the SSC afterglow model either with and without energy injection is considered. Therefore, in addition to the work presented by \cite{2019ApJ...883..134T}, the CRs of SSC afterglow models are needed to explain those bursts in the 2FLGC that cannot be interpreted in the synchrotron scenario (e.g., those with photon energies above 10 GeV).

\section{acknowledgements}
We would like to express our gratitude to the anonymous referee for their careful reading of the manuscript and insightful recommendations that helped improve quality and clarity of this manuscript.  NF acknowledges financial support  from UNAM-DGAPA-PAPIIT  through  grant IN106521. 

%\end{acknowledgements}

\bibliography{references}

\begin{thebibliography}{}
\expandafter\ifx\csname natexlab\endcsname\relax\def\natexlab#1{#1}\fi
\providecommand{\url}[1]{\href{#1}{#1}}
\providecommand{\dodoi}[1]{doi:~\href{http://doi.org/#1}{\nolinkurl{#1}}}
\providecommand{\doeprint}[1]{\href{http://ascl.net/#1}{\nolinkurl{http://ascl.net/#1}}}
\providecommand{\doarXiv}[1]{\href{https://arxiv.org/abs/#1}{\nolinkurl{https://arxiv.org/abs/#1}}}

\bibitem[{{Aartsen} {et~al.}(2015){Aartsen}, {Ackermann}, {Adams}, {Aguilar},
  {Ahlers}, {Ahrens}, {Altmann}, {Anderson}, {Arguelles}, {Arlen}, \&
  et~al.}]{2015ApJ...805L...5A}
{Aartsen}, M.~G., {Ackermann}, M., {Adams}, J., {et~al.} 2015, \apjl, 805, L5,
  \dodoi{10.1088/2041-8205/805/1/L5}

\bibitem[{{Aartsen} {et~al.}(2016){Aartsen}, {Abraham}, {Ackermann}, {Adams},
  {Aguilar}, {Ahlers}, {Ahrens}, {Altmann}, {Anderson}, {Ansseau}, \&
  et~al.}]{2016ApJ...824..115A}
{Aartsen}, M.~G., {Abraham}, K., {Ackermann}, M., {et~al.} 2016, \apj, 824,
  115, \dodoi{10.3847/0004-637X/824/2/115}

\bibitem[{{Abbasi} {et~al.}(2012){Abbasi}, {Abdou}, {Abu-Zayyad}, {Ackermann},
  {Adams}, {Aguilar}, {Ahlers}, {Altmann}, {Andeen}, {Auffenberg}, \&
  et~al.}]{2012Natur.484..351A}
{Abbasi}, R., {Abdou}, Y., {Abu-Zayyad}, T., {et~al.} 2012, \nat, 484, 351,
  \dodoi{10.1038/nature11068}

\bibitem[{{Ackermann} \& et~al.(2010)}]{2010ApJ...716.1178A}
{Ackermann}, M., \& et~al. 2010, \apj, 716, 1178,
  \dodoi{10.1088/0004-637X/716/2/1178}

\bibitem[{{Ackermann} \& et~al.(2013)}]{2013ApJ...763...71A}
---. 2013, \apj, 763, 71, \dodoi{10.1088/0004-637X/763/2/71}

\bibitem[{{Ajello} {et~al.}(2019){Ajello}, {Arimoto}, {Axelsson}, {Baldini},
  {Barbiellini}, \& {et al.}}]{Ajello_2019}
{Ajello}, M., {Arimoto}, M., {Axelsson}, M., {et~al.} 2019, \apj, 878, 52,
  \dodoi{10.3847/1538-4357/ab1d4e}

\bibitem[{{Band} {et~al.}(1993){Band}, {Matteson}, {Ford}, {Schaefer},
  {Palmer}, {Teegarden}, {Cline}, {Briggs}, {Paciesas}, {Pendleton}, {Fishman},
  {Kouveliotou}, {Meegan}, {Wilson}, \& {Lestrade}}]{1993ApJ...413..281B}
{Band}, D., {Matteson}, J., {Ford}, L., {et~al.} 1993, \apj, 413, 281,
  \dodoi{10.1086/172995}

\bibitem[{{Barthelmy} {et~al.}(2005){Barthelmy}, {Cannizzo}, {Gehrels},
  {Cusumano}, {Mangano}, \& {et al.}}]{2005ApJ...635L.133B}
{Barthelmy}, S.~D., {Cannizzo}, J.~K., {Gehrels}, N., {et~al.} 2005, \apjl,
  635, L133, \dodoi{10.1086/499432}

\bibitem[{{Becerra} {et~al.}(2019){Becerra}, {Watson}, {Fraija}, {Butler},
  {Lee}, \& {et al.}}]{2019ApJ...872..118B}
{Becerra}, R.~L., {Watson}, A.~M., {Fraija}, N., {et~al.} 2019, \apj, 872, 118,
  \dodoi{10.3847/1538-4357/ab0026}

\bibitem[{{B{\"o}ttcher} \& {Dermer}(2000)}]{2000ApJ...532..281B}
{B{\"o}ttcher}, M., \& {Dermer}, C.~D. 2000, \apj, 532, 281,
  \dodoi{10.1086/308580}

\bibitem[{{Burrows} {et~al.}(2005){Burrows}, {Romano}, {Falcone}, {Kobayashi},
  {Zhang}, \& {et al.}}]{2005Sci...309.1833B}
{Burrows}, D.~N., {Romano}, P., {Falcone}, A., {et~al.} 2005, Science, 309,
  1833, \dodoi{10.1126/science.1116168}

\bibitem[{{Chevalier} \& {Li}(2000)}]{2000ApJ...536..195C}
{Chevalier}, R.~A., \& {Li}, Z.-Y. 2000, \apj, 536, 195, \dodoi{10.1086/308914}

\bibitem[{{Chincarini} {et~al.}(2007){Chincarini}, {Moretti}, {Romano},
  {Falcone}, {Morris}, \& {et al.}}]{2007ApJ...671.1903C}
{Chincarini}, G., {Moretti}, A., {Romano}, P., {et~al.} 2007, \apj, 671, 1903,
  \dodoi{10.1086/521591}

\bibitem[{{Dai} \& {Lu}(1998)}]{1998MNRAS.298...87D}
{Dai}, Z.~G., \& {Lu}, T. 1998, \mnras, 298, 87,
  \dodoi{10.1046/j.1365-8711.1998.01681.x}

\bibitem[{{Dai} {et~al.}(2006){Dai}, {Wang}, {Wu}, \&
  {Zhang}}]{2006Sci...311.1127D}
{Dai}, Z.~G., {Wang}, X.~Y., {Wu}, X.~F., \& {Zhang}, B. 2006, Science, 311,
  1127, \dodoi{10.1126/science.1123606}

\bibitem[{{Dainotti} {et~al.}(2021{\natexlab{a}}){Dainotti}, {Lenart},
  {Fraija}, {Nagataki}, {Warren}, {De Simone}, {Srinivasaragavan}, \&
  {Mata}}]{2021PASJ...73..970D}
{Dainotti}, M.~G., {Lenart}, A.~{\L}., {Fraija}, N., {et~al.}
  2021{\natexlab{a}}, \pasj, 73, 970, \dodoi{10.1093/pasj/psab057}

\bibitem[{{Dainotti} {et~al.}(2021{\natexlab{b}}){Dainotti}, {Omodei},
  {Srinivasaragavan}, {Vianello}, {Willingale}, {O'Brien}, {Nagataki},
  {Petrosian}, {Nuygen}, {Hernandez}, {Axelsson}, {Bissaldi}, \&
  {Longo}}]{2021ApJS..255...13D}
{Dainotti}, M.~G., {Omodei}, N., {Srinivasaragavan}, G.~P., {et~al.}
  2021{\natexlab{b}}, \apjs, 255, 13, \dodoi{10.3847/1538-4365/abfe17}

\bibitem[{{Dall'Osso} {et~al.}(2017){Dall'Osso}, {Perna}, {Tanaka}, \&
  {Margutti}}]{2017MNRAS.464.4399D}
{Dall'Osso}, S., {Perna}, R., {Tanaka}, T.~L., \& {Margutti}, R. 2017, \mnras,
  464, 4399, \dodoi{10.1093/mnras/stw2695}

\bibitem[{{Duncan} \& {Thompson}(1992)}]{1992ApJ...392L...9D}
{Duncan}, R.~C., \& {Thompson}, C. 1992, \apjl, 392, L9, \dodoi{10.1086/186413}

\bibitem[{{Fraija}(2015)}]{2015ApJ...804..105F}
{Fraija}, N. 2015, \apj, 804, 105, \dodoi{10.1088/0004-637X/804/2/105}

\bibitem[{{Fraija} {et~al.}(2019{\natexlab{a}}){Fraija}, {Barniol Duran},
  {Dichiara}, \& {Beniamini}}]{2019ApJ...883..162F}
{Fraija}, N., {Barniol Duran}, R., {Dichiara}, S., \& {Beniamini}, P.
  2019{\natexlab{a}}, \apj, 883, 162, \dodoi{10.3847/1538-4357/ab3ec4}

\bibitem[{{Fraija} {et~al.}(2019{\natexlab{b}}){Fraija}, {Dichiara},
  {Pedreira}, {Galvan-Gamez}, {Becerra}, {Barniol Duran}, \&
  {Zhang}}]{2019ApJ...879L..26F}
{Fraija}, N., {Dichiara}, S., {Pedreira}, A.~C. C. d. E.~S., {et~al.}
  2019{\natexlab{b}}, \apjl, 879, L26, \dodoi{10.3847/2041-8213/ab2ae4}

\bibitem[{{Fraija} {et~al.}(2020){Fraija}, {Laskar}, {Dichiara}, {Beniamini},
  {Duran}, {Dainotti}, \& {Becerra}}]{2020ApJ...905..112F}
{Fraija}, N., {Laskar}, T., {Dichiara}, S., {et~al.} 2020, \apj, 905, 112,
  \dodoi{10.3847/1538-4357/abc41a}

\bibitem[{{Fraija} {et~al.}(2016{\natexlab{a}}){Fraija}, {Lee}, \&
  {Veres}}]{2016ApJ...818..190F}
{Fraija}, N., {Lee}, W., \& {Veres}, P. 2016{\natexlab{a}}, \apj, 818, 190,
  \dodoi{10.3847/0004-637X/818/2/190}

\bibitem[{{Fraija} {et~al.}(2017{\natexlab{a}}){Fraija}, {Lee}, {Araya},
  {Veres}, {Barniol Duran}, \& {Guiriec}}]{2017ApJ...848...94F}
{Fraija}, N., {Lee}, W.~H., {Araya}, M., {et~al.} 2017{\natexlab{a}}, \apj,
  848, 94, \dodoi{10.3847/1538-4357/aa8d65}

\bibitem[{{Fraija} {et~al.}(2016{\natexlab{b}}){Fraija}, {Lee}, {Veres}, \&
  {Barniol Duran}}]{2016ApJ...831...22F}
{Fraija}, N., {Lee}, W.~H., {Veres}, P., \& {Barniol Duran}, R.
  2016{\natexlab{b}}, \apj, 831, 22, \dodoi{10.3847/0004-637X/831/1/22}

\bibitem[{{Fraija} {et~al.}(2021){Fraija}, {Veres}, {Beniamini},
  {Galvan-Gamez}, {Metzger}, {Barniol Duran}, \&
  {Becerra}}]{2021ApJ...918...12F}
{Fraija}, N., {Veres}, P., {Beniamini}, P., {et~al.} 2021, \apj, 918, 12,
  \dodoi{10.3847/1538-4357/ac0aed}

\bibitem[{{Fraija} {et~al.}(2017{\natexlab{b}}){Fraija}, {Veres}, {Zhang},
  {Barniol Duran}, {Becerra}, {Zhang}, {Lee}, {Watson}, {Ordaz-Salazar}, \&
  {Galvan-Gamez}}]{2017ApJ...848...15F}
{Fraija}, N., {Veres}, P., {Zhang}, B.~B., {et~al.} 2017{\natexlab{b}}, \apj,
  848, 15, \dodoi{10.3847/1538-4357/aa8a72}

\bibitem[{{Fraija} {et~al.}(2019{\natexlab{c}}){Fraija}, {Dichiara},
  {Pedreira}, {Galvan-Gamez}, {Becerra}, {Montalvo}, {Montero}, {Betancourt
  Kamenetskaia}, \& {Zhang}}]{2019ApJ...885...29F}
{Fraija}, N., {Dichiara}, S., {Pedreira}, A.~C. C. d. E.~S., {et~al.}
  2019{\natexlab{c}}, \apj, 885, 29, \dodoi{10.3847/1538-4357/ab3e4b}

\bibitem[{{Ghisellini} {et~al.}(2010){Ghisellini}, {Ghirlanda}, {Nava}, \&
  {Celotti}}]{2010MNRAS.403..926G}
{Ghisellini}, G., {Ghirlanda}, G., {Nava}, L., \& {Celotti}, A. 2010, \mnras,
  403, 926, \dodoi{10.1111/j.1365-2966.2009.16171.x}

\bibitem[{{He} {et~al.}(2011){He}, {Wu}, {Toma}, {Wang}, \&
  {M{\'e}sz{\'a}ros}}]{2011ApJ...733...22H}
{He}, H.-N., {Wu}, X.-F., {Toma}, K., {Wang}, X.-Y., \& {M{\'e}sz{\'a}ros}, P.
  2011, \apj, 733, 22, \dodoi{10.1088/0004-637X/733/1/22}

\bibitem[{{King} {et~al.}(2005){King}, {O'Brien}, {Goad}, {Osborne}, {Olsson},
  \& {Page}}]{2005ApJ...630L.113K}
{King}, A., {O'Brien}, P.~T., {Goad}, M.~R., {et~al.} 2005, \apjl, 630, L113,
  \dodoi{10.1086/496881}

\bibitem[{{Kouveliotou} {et~al.}(1993){Kouveliotou}, {Meegan}, {Fishman},
  {Bhat}, {Briggs}, {Koshut}, {Paciesas}, \& {Pendleton}}]{1993ApJ...413L.101K}
{Kouveliotou}, C., {Meegan}, C.~A., {Fishman}, G.~J., {et~al.} 1993, \apjl,
  413, L101, \dodoi{10.1086/186969}

\bibitem[{{Kouveliotou} {et~al.}(2013){Kouveliotou}, {Granot}, {Racusin},
  {Bellm}, {Vianello}, {Oates}, {Fryer}, {Boggs}, {Christensen}, {Craig},
  {Dermer}, {Gehrels}, {Hailey}, {Harrison}, {Melandri}, {McEnery}, {Mundell},
  {Stern}, {Tagliaferri}, \& {Zhang}}]{2013ApJ...779L...1K}
{Kouveliotou}, C., {Granot}, J., {Racusin}, J.~L., {et~al.} 2013, \apjl, 779,
  L1, \dodoi{10.1088/2041-8205/779/1/L1}

\bibitem[{{Kumar} \& {Barniol Duran}(2009)}]{2009MNRAS.400L..75K}
{Kumar}, P., \& {Barniol Duran}, R. 2009, \mnras, 400, L75,
  \dodoi{10.1111/j.1745-3933.2009.00766.x}

\bibitem[{{Kumar} \& {Barniol Duran}(2010)}]{2010MNRAS.409..226K}
---. 2010, \mnras, 409, 226, \dodoi{10.1111/j.1365-2966.2010.17274.x}

\bibitem[{{Kumar} \& {Zhang}(2015)}]{2015PhR...561....1K}
{Kumar}, P., \& {Zhang}, B. 2015, \physrep, 561, 1,
  \dodoi{10.1016/j.physrep.2014.09.008}

\bibitem[{{Li} {et~al.}(2002){Li}, {Dai}, \& {Lu}}]{2002MNRAS.330..955L}
{Li}, Z., {Dai}, Z.~G., \& {Lu}, T. 2002, \mnras, 330, 955,
  \dodoi{10.1046/j.1365-8711.2002.05141.x}

\bibitem[{{M{\'e}sz{\'a}ros} \& {Rees}(1997)}]{1997ApJ...476..232M}
{M{\'e}sz{\'a}ros}, P., \& {Rees}, M.~J. 1997, \apj, 476, 232

\bibitem[{{Metzger} {et~al.}(2011){Metzger}, {Giannios}, {Thompson},
  {Bucciantini}, \& {Quataert}}]{2011MNRAS.413.2031M}
{Metzger}, B.~D., {Giannios}, D., {Thompson}, T.~A., {Bucciantini}, N., \&
  {Quataert}, E. 2011, \mnras, 413, 2031,
  \dodoi{10.1111/j.1365-2966.2011.18280.x}

\bibitem[{{Moderski} {et~al.}(2000){Moderski}, {Sikora}, \&
  {Bulik}}]{2000ApJ...529..151M}
{Moderski}, R., {Sikora}, M., \& {Bulik}, T. 2000, \apj, 529, 151,
  \dodoi{10.1086/308257}

\bibitem[{{Nakar} {et~al.}(2009){Nakar}, {Ando}, \&
  {Sari}}]{2009ApJ...703..675N}
{Nakar}, E., {Ando}, S., \& {Sari}, R. 2009, \apj, 703, 675,
  \dodoi{10.1088/0004-637X/703/1/675}

\bibitem[{{Paczy{\'n}ski}(1998)}]{1998ApJ...494L..45P}
{Paczy{\'n}ski}, B. 1998, \apjl, 494, L45, \dodoi{10.1086/311148}

\bibitem[{{Panaitescu}(2019)}]{2019ApJ...886..106P}
{Panaitescu}, A. 2019, \apj, 886, 106, \dodoi{10.3847/1538-4357/ab4e17}

\bibitem[{{Perna} {et~al.}(2006){Perna}, {Armitage}, \&
  {Zhang}}]{2006ApJ...636L..29P}
{Perna}, R., {Armitage}, P.~J., \& {Zhang}, B. 2006, \apjl, 636, L29,
  \dodoi{10.1086/499775}

\bibitem[{{Piran}(1999)}]{1999PhR...314..575P}
{Piran}, T. 1999, \physrep, 314, 575, \dodoi{10.1016/S0370-1573(98)00127-6}

\bibitem[{{Proga} \& {Zhang}(2006)}]{2006MNRAS.370L..61P}
{Proga}, D., \& {Zhang}, B. 2006, \mnras, 370, L61,
  \dodoi{10.1111/j.1745-3933.2006.00189.x}

\bibitem[{{Santana} {et~al.}(2014){Santana}, {Barniol Duran}, \&
  {Kumar}}]{2014ApJ...785...29S}
{Santana}, R., {Barniol Duran}, R., \& {Kumar}, P. 2014, \apj, 785, 29,
  \dodoi{10.1088/0004-637X/785/1/29}

\bibitem[{{Sari} \& {Esin}(2001)}]{2001ApJ...548..787S}
{Sari}, R., \& {Esin}, A.~A. 2001, \apj, 548, 787, \dodoi{10.1086/319003}

\bibitem[{{Sari} {et~al.}(1998){Sari}, {Piran}, \&
  {Narayan}}]{1998ApJ...497L..17S}
{Sari}, R., {Piran}, T., \& {Narayan}, R. 1998, \apjl, 497, L17,
  \dodoi{10.1086/311269}

\bibitem[{{Srinivasaragavan} {et~al.}(2020){Srinivasaragavan}, {Dainotti},
  {Fraija}, {Hernandez}, {Nagataki}, {Lenart}, {Bowden}, \&
  {Wagner}}]{2020ApJ...903...18S}
{Srinivasaragavan}, G.~P., {Dainotti}, M.~G., {Fraija}, N., {et~al.} 2020,
  \apj, 903, 18, \dodoi{10.3847/1538-4357/abb702}

\bibitem[{{Tak} {et~al.}(2019){Tak}, {Omodei}, {Uhm}, {Racusin}, {Asano}, \&
  {McEnery}}]{2019ApJ...883..134T}
{Tak}, D., {Omodei}, N., {Uhm}, Z.~L., {et~al.} 2019, \apj, 883, 134,
  \dodoi{10.3847/1538-4357/ab3982}

\bibitem[{{Thompson}(1994)}]{1994MNRAS.270..480T}
{Thompson}, C. 1994, \mnras, 270, 480, \dodoi{10.1093/mnras/270.3.480}

\bibitem[{{Usov}(1992)}]{1992Natur.357..472U}
{Usov}, V.~V. 1992, \nat, 357, 472, \dodoi{10.1038/357472a0}

\bibitem[{{Wang} {et~al.}(2010){Wang}, {He}, {Li}, {Wu}, \&
  {Dai}}]{2010ApJ...712.1232W}
{Wang}, X.-Y., {He}, H.-N., {Li}, Z., {Wu}, X.-F., \& {Dai}, Z.-G. 2010, \apj,
  712, 1232, \dodoi{10.1088/0004-637X/712/2/1232}

\bibitem[{{Waxman} \& {Bahcall}(1997)}]{1997PhRvL..78.2292W}
{Waxman}, E., \& {Bahcall}, J. 1997, \prl, 78, 2292,
  \dodoi{10.1103/PhysRevLett.78.2292}

\bibitem[{{Woosley}(1993)}]{1993ApJ...405..273W}
{Woosley}, S.~E. 1993, \apj, 405, 273, \dodoi{10.1086/172359}

\bibitem[{{Wu} {et~al.}(2005){Wu}, {Dai}, {Huang}, \&
  {Lu}}]{2005ApJ...619..968W}
{Wu}, X.~F., {Dai}, Z.~G., {Huang}, Y.~F., \& {Lu}, T. 2005, \apj, 619, 968,
  \dodoi{10.1086/426666}

\bibitem[{{Zhang}(2019)}]{2019arXiv191109862Z}
{Zhang}, B. 2019, arXiv e-prints, arXiv:1911.09862.
\newblock \doarXiv{1911.09862}

\bibitem[{{Zhang} {et~al.}(2006){Zhang}, {Fan}, {Dyks}, {Kobayashi},
  {M{\'e}sz{\'a}ros}, {Burrows}, {Nousek}, \& {Gehrels}}]{2006ApJ...642..354Z}
{Zhang}, B., {Fan}, Y.~Z., {Dyks}, J., {et~al.} 2006, \apj, 642, 354,
  \dodoi{10.1086/500723}

\end{thebibliography}
\bibliographystyle{aasjournal}

\addcontentsline{toc}{chapter}{Bibliography}

\begin{comment}
\begin{figure*}
\centering
\includegraphics[width=1\textwidth]{SSC_Closure_Relations_v2}
\caption{CRs of SSC forward-shock model evolving in stellar-wind (left) and constant-density (right) medium with  spectral and temporal indexes reported in the Second Fermi-LAT GRB catalog}
\label{lat_2catalog}
\end{figure*}
\end{comment}

\end{document}